# Droplets with circular stagnation lines: combined effects of viscoelastic and inertial forces on drop shapes

A. Emamian[1] M. Norouzi[1] and M. Davoodi[2,3] *

[1] *Faculty of Mechanical Engineering, Shahrood University of Technology, Shahrood, Iran.*
[2] *School of Engineering, University of Liverpool, Brownlow Hill, Liverpool, L69 3GH, UK.*
[3] *Schlumberger Cambridge Research, High Cross, Madingley Road, Cambridge, CB3 0EL, UK*

**Abstract**

Hydrodynamic problems with stagnation points are of particular importance in fluid mechanics as they allow study and investigation of elongational flows. In this article, the uniaxial elongational flow appearing at the surface of a viscoelastic drop and its role on the deformation of the droplet at low inertial regimes is studied. In studies related to viscoelastic droplets falling/raising in an immiscible Newtonian fluids, it is well known that by increasing the Deborah number (the ratio of the relaxation time of the interior fluid to a reference time scale) the droplet might lose its sphericity and obtain a dimple at the rear end. In this work, the drop deformation is investigated in detail to study the reason behind this transformation. We will show that as the contribution of elastic and inertial forces are increased, the stagnation points at the rear and front sides of the droplet are expanded to create a region of elongational dominated flows. At this stage, due to a combined effect of the shear thickening behavior of the elongational viscosity in viscoelastic fluids and the contribution of the inertial force, the interior phase is squeezed and consequently the droplet finds a shape similar to an oblate. As these non-linear forces are increased further, an additional *circular stagnation line* appears on the droplet surface in the external field, pulling the droplet surface outward and therefore creating a dimple shape at the rear end. Furthermore, the influence of inertia and viscoelastic properties are also studied on the motion, the drag coefficient and terminal velocity of drops.

---

* Email address for correspondence: mdavoodi@slb.com

## 1. Introduction

Extensional viscosity (also known as elongational viscosity) is a viscosity coefficient measuring the resistance of the fluid to the extensional deformation of the flow. For Newtonian fluids, it is known that its value is Trouton ratio of the shear viscosity [1], while for polymeric liquids this coefficient is a thickening function of the applied deformation rate, i.e. the strain rate, and can be dramatically larger than its equivalent value in Newtonian fluids [2]. To obtain and study this parameter, geometries are designed in such a way that the fluid is subjected to a homogenous elongational flow for a sufficiently long time, so a fluid element can reach a steady state in the Lagrangian frame. Such flow conditions may be observed as fluid passes through intersections such as T- shaped [3] or cross-slot [4] geometries or an obstacle [4, 5]. Among these the cross-slot geometry is probably the most common flow geometry that has been utilized for investigation of extensional flow fields [6, 7]. In Newtonian fluids, due to presence of opposing inlets and outlets in this geometry, a flow field with a *single* strand of elongational flow at the inlet and outlet arms appears, producing a stagnation point at the centre of the geometry. In the research carried out by Davoodi et al. [5], the modifying effect of viscoelasticity on the flow field was investigated, showing the existence of **three** strands of elongational-dominated fields, instead of a *single* strand as found in Newtonian flows, downstream of the "free" stagnation point. This study was carried out for single phase fluid flows, highlighting that as the Deborah number is increased the elastic modification on the flow field downstream of the stagnation point might be substantial, generating three high normal-stress regions. Recently, Neithammer et al. [8] have shown that once a bubble is rising in a viscoelastic fluid, the normal stress distribution at the vicinity of the rear end stagnation point region will expand and above a critical Deborah number, three strands of normal stress dominated flows will be appeared (in 2D presentation) which was coincided with the appearance of cusp like shape in the bubble and a jump of terminal velocity in the study.

Besides the interest arising from investigation of elongational dominated flows, drop studies are of great importance [9] in removal of suspended and colloidal impurities in water-treatment industries. During flocculation process, the sedimentation rate and so the drag coefficient depends on various parameters such as droplet size and rheological properties of the viscoelastic material. Furthermore, such fundamental studies find crucial roles in processes such as coating [10], spray painting [11], spray cooling [12], blending molten polymers [13, 14], ink-jet printers [15, 16], as well as the use of immiscible drops in direct-contact heat exchangers [17, 18] and nearly miscible drops in liquid-liquid extraction processes [19]. In the recent decades, various studies and analyses have been carried out for falling or rising liquid drops through another immiscible fluid, indicating the importance of this issue. In such problems a jump of normal forces appears at the interface of two fluids which is balanced with the interfacial tension force as [20, 21]:

$$\boldsymbol{n}.(\boldsymbol{\tau}_e - \boldsymbol{\tau}_i).\boldsymbol{n} - p_e + p_i = \frac{1}{Ca}\left(\frac{1}{R_1} + \frac{1}{R_2}\right), \tag{1}$$

Where $\boldsymbol{\tau}$ and $p$ are stress and pressure, index "$e$" and "$i$" are referring to exterior and interior fluids, $n$ is the unit vector normal to the surface of the droplet, $Ca$ is the Capillary number (the ratio of viscous to interfacial forces) and $R_1$ and $R_2$ are the principal radii of interface curvature. Newtonian creeping motion of liquid drops was initially addressed in the seminal analytical studies of Hadamard [20] and Rybczynski [21]. In those works, it has been shown that for creeping Newtonian drops, the shape remains exactly spherical. In such problems, as a result of a linear correlations in the governing equations, the distribution of normal forces at the interface of two fluids stays uniform, i.e. the terms on the left-hand side of equation (1) are equal to a constant, so the interfacial tension is able to balance this constant value with a constant curvature (the droplet finds a spherical shape). Taylor and Acrivos [22] extended this analysis to include the nonlinear effects of the inertial forces using a perturbation analysis, showing that as the Reynolds number increases the droplet loses its sphericity and will take an oblate shape. In this work it has been shown that unlike what is observed in the creeping Newtonian regime, the jump on the normal forces across the interface is non-uniform across the droplet surface (the left-hand side of equation (1) will be a function $\theta$ variable in the employed spherical coordinate system), and so the curvature of interface will be a function of $\theta$ variable, i.e. $R_1(\theta)$ and $R_2(\theta)$.

Non-Newtonian materials are frequently used in a variety of technological applications, ranging from biotechnology to mechanical, civil and chemical engineering [23]. In order to achieve an appropriate design in such processes, a reasonable understanding of the effect of drop size, shape, deformation and terminal velocity on its motion is crucial [24]. Sostarecz and Belmonte [25] studied experimentally and analytically the falling of creeping viscoelastic immiscible drops in Newtonian media. In this study it has been shown that with an increase in the drop volume and in the absence of any inertial force, the droplet loses its sphericity and obtains an oblate shape. A further increase in volume of droplet then causes the appearance of a dimple at the rear end of the drop. In such viscoelastic problems, the non-linear properties of viscoelastic fluids lead to a non-uniform distribution of normal forces across the interface of two fluids and consequently results in a non-spherical shape. Recently, a similar analytical technique has been used to extend the analysis to investigate the thermocapillary motion of Newtonian droplets in viscoelastic fluids [26].

In the current study, the motion of viscoelastic drops falling through Newtonian fluids at low Reynolds numbers is investigated using both experimental and analytical approaches. Here, the Giesekus model is used as the constitutive equation and a triple perturbation technique is employed to solve the governing equations. Non-dimensional Reynolds, Deborah and capillary numbers are used as perturbation parameters. The analytical results are then validated using our experimental data for the shape and motion of nonlinear viscoelastic drops to include small but non-negligible inertia and elastic effects. In the previous studies, the deformation of droplets has been simply attributed the non-linear contributions of inertia and elastic forces but the physics behind the shape change was not explained/investigated in detail. In the current work, it is shown that as *De* or *Re* increases, besides the stagnation point at the rear end and the front side of the droplet a circular stagnation line starts to expand from this point, which is believed to be responsible for the complex shape transformation of the droplets.

**Experimental approach**

In this study, viscoelastic drops are created using four different polymeric solutions. A mixture of deionized water and glycerin solution (20:80 ratio by volume water to glycerol) is used as the Newtonian solvent. This solvent is made viscoelastic by adding 200, 300, 400 and 600 ppm polyacrylamide (PAM). For simplicity, hereafter these four viscoelastic solutions are named PAM-1, PAM-2, PAM-3 and PAM-4, respectively. The degree of purity of glycerin is 99% and the molecular weight of polyacrylamide is $5\times10^6$ g/mol. Polyacrylamide is a commercially relevant [27] cationic polymer utilized mainly for water treatment due to its high efficiency and rapid dissolution. These solutions are employed as flocculants in the removal of suspended particles from sewage and industrial effluents (e.g., wastewater from paper mills). Water and polyacrylamide are poured in a magnetic hotplate stirrer, and they are mixed for 24 hours at 80 rpm and 25°C. Then, glycerin is added to the mixture, and the fluid is stirred for 24 hours to produce a homogenous solution. After one day of mixing, the fluid becomes homogeneous [28]. The density of the solution is estimated to be $\rho_i^* = 1.2 \; \frac{g}{cm^3}$ (here "i" refers to the interior fluid). Note that the density has not much changed by the addition of the polymer. In this article all dimensional parameters are marked with an asterisk. Silicon oil with a viscosity of 600 mPa.s is used as the exterior fluid as it is immiscible with water. The interfacial tension between the two fluids is measured via the pendant drop technique using an OCA20 Data Physics Instrument (presented in Table 1).

A rectangular Plexiglas tank with 15×15 cm² cross-section and 50 cm height is filled with silicon oil. Some specific volumes (the width of cross-section is 10 times of the largest drop radius used in the experiments) of a viscoelastic fluid are dropped from a needle attached to a burette. As shown in Fig. 1, the tip of the needle is placed inside the silicon oil so that the drop falls into the bulk fluid. During all experiments, the temperature is kept constant at 25°C. The motion of the falling drop is recorded using a high-speed camera (pco. dimax S) and the images are taken with the frame rate, the exposure time and the resolution of 10000 fps, 0.0001 s, and 850×750 pixels, respectively. Appropriate lighting is required for high quality imaging and thus two projectors (UNIMAT) are used for backlighting. Fig. 1 shows a schematic view of the experimental setup.

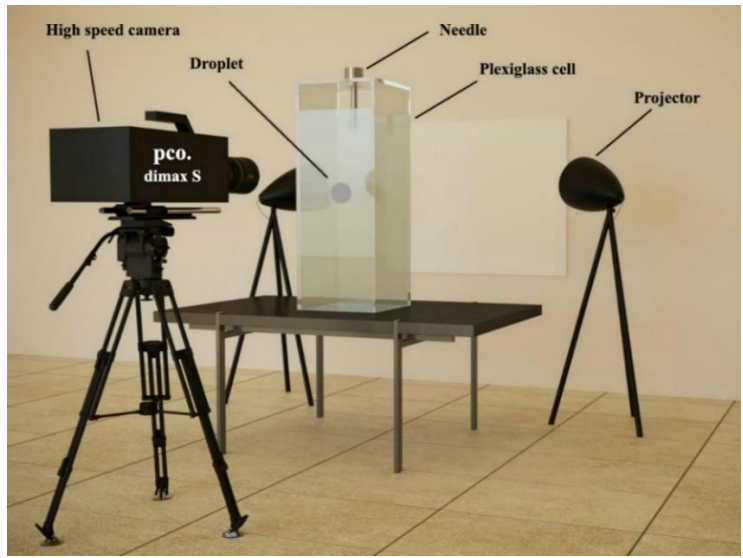

Fig. 1. Schematic sketch of the experimental setup.

Analyzing the storage modulus ($G'$) and the loss modulus ($G''$) is a useful tool to estimate relaxation times of these fluids. Here, a rheometer (MCR-302, Anton Paar Co.) is used to measure the rheological properties. Both the amplitude and frequency sweep tests are carried out at 25°C, via a cone and plate geometry with diameter of 50 mm (CP50). The amplitude sweep test is performed at the constant frequency of 1 Hz to identify the range of the strain amplitude for which the oscillatory test data are linear. According to the results, both $G'$ and $G''$ of the liquids are constant up to 40% of the strain and the test is therefore linear in this range. A fixed strain amplitude of 10% is used for the frequency sweep test of the viscoelastic liquids. Fig. 2 shows the results of the frequency sweep test in the range of 0.02 to 10 rad/s. The generalized Maxwell model is used to estimate the average relaxation time. The response of the generalized Maxwell model for the oscillatory test is as follows:

$$G' = \sum_{j=1}^{n} \frac{\eta_j^* \lambda_j^* \omega^{*2}}{1+\lambda_j^{*2} \omega^{*2}}, \tag{2a}$$

$$G'' = \eta_s^* \omega^* + \sum_{j=1}^{n} \frac{\eta_j^* \omega^*}{1+\lambda_j^{*2} \omega^{*2}}, \tag{2b}$$

$$\overline{\lambda^*} = \frac{\sum_{j=1}^{n} \eta_j^* \lambda_j^*}{\sum_{j=1}^{n} \eta_j^*}. \tag{2c}$$

The hybrid genetic algorithm method is used to fit the response of the generalized Maxwell model (Eqs. (2a) and (2b)) on the experimental data of $G'$ and $G''$. In other words, the material parameters of the model ($\eta_j^*$ and $\lambda_j^*$) are determined by minimizing the deviation between the model and experimental data of the frequency sweep test. In Fig. 2 the fit and the experimental data are presented. Using Eq. (2c), the average relaxation time of viscoelastic liquids could be estimated (presented in Table 1).

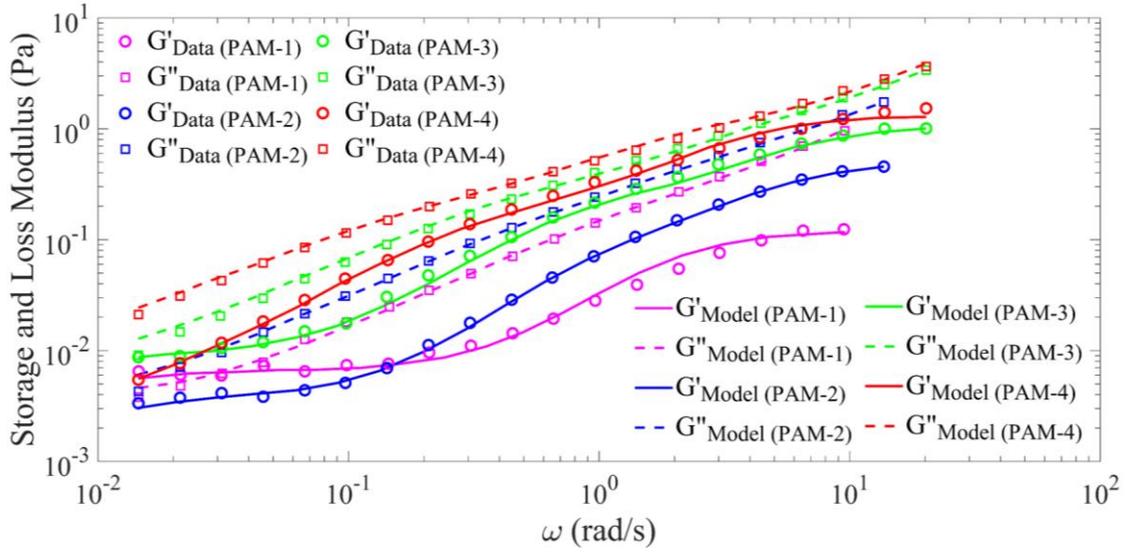

Fig. 2. The storage and loss moduli ($G'$ and $G''$) versus the frequency for four viscoelastic liquids at 25°C.

In this paper, the Giesekus model is used as the constitutive equation to analyze the dynamics of falling polymeric drops at law Reynolds numbers. This model is derived based on the concept of a configuration-dependent molecular mobility and it is more popular than retarded motion expansion models owing to a fewer constants (as a four-constant model). The constants of this model are well-defined based on the viscoelastic material modulus. The Giesekus model is considered as a nonlinear constitutive equation and its results have been confirmed for numerous viscoelastic flows. The Giesekus model is defined as follows:

$$\boldsymbol{\tau}^* = \boldsymbol{\tau}_p^* + \boldsymbol{\tau}_s^*, \tag{3a}$$

$$\boldsymbol{\tau}_p^* + \lambda^* \hat{d}\boldsymbol{\tau}_p^* + \frac{\alpha \lambda^*}{\eta_{p,0}^*}\left(\boldsymbol{\tau}_p^* \cdot \boldsymbol{\tau}_p^*\right) = 2\eta_{p,0}^* \boldsymbol{D}^*, \tag{3b}$$

$$\boldsymbol{\tau}_s^* = 2\eta_s^* \boldsymbol{D}^*. \tag{3c}$$

In this presentation, the model is defined for a viscoelastic solution for which the stress is a sum of the stresses of the Newtonian solvent, $\boldsymbol{\tau}_s^*$, and the polymeric additive, $\boldsymbol{\tau}_p^*$. Here, $\eta_s^*$ and $\eta_{p,0}^*$ are the viscosity of the Newtonian solvent and the polymeric contribution as shear rate tends to zero, respectively. In addition, $\alpha$ is the model mobility factor and its origin can be associated with anisotropic Brownian motion and/or anisotropic hydrodynamic drag on the constitutive polymer molecules [29]. The mobility factor indicates the level of nonlinearity of the model and, although it can vary from zero to one from the molecular theory point of view, the experimental observations reveal that $\alpha$ is between zero to 0.5 for viscoelastic materials known in nature [2]. The operator $\hat{d}$ is the upper convective derivative which is defined as follows:

$$\hat{d}(A) = \left(\frac{\partial}{\partial t^*} + u^* \cdot \nabla\right)(A) - \left((\nabla u^{*T})(A) + (A)(\nabla u^*)\right), \tag{4}$$

where $A$ is an arbitrary second order tensor. The Giesekus equation can model the rheological response of concentrated viscoelastic solutions and polymeric melts, in addition to dilute polymeric solutions [29]. For the limiting case of $\alpha = 0$, the model is reduced to the Oldroyd-B constitutive equation (i.e., a quasi-linear model), which is suitable for dilute polymeric solutions and Boger fluids [30]. Once $\lambda^* = 0$, the Giesekus constitutive equation is reduced to the Newtonian model.

The zero shear-rate viscosity and the mobility factor of the Giesekus model can be obtained by curve fitting using rheological data (viscometric tests). The response of the Giesekus model for the steady shear test is as follows [29, 31]:

$$\frac{\eta^*}{\eta_0^*} = \frac{(1-f)^2}{1+(1-2\alpha)f}, \tag{5}$$

$$f = \frac{1-\chi}{1+(1-2\alpha)\chi}, \chi^2 = \frac{\left[1+16\alpha(1-\alpha)(\lambda^*\dot{\gamma}^*)^2\right]^{\frac{1}{2}}-1}{8\alpha(1-\alpha)(\lambda^*\dot{\gamma}^*)^2}. \tag{6}$$

Here, a hybrid genetic algorithm is used to minimize the difference between the response of the Giesekus model (Eqs. (5) and (6)) and viscometric tests performed using the rheometer with a cylindrical geometry. Fig. 3 depicts the viscosity curves of our polymeric solutions and the fitted response of the Giesekus model for the steady shear test. The extracted data of the curved fitting of the Giesekus model for both the frequency sweep tests and the viscometric tests are listed in Table 1.

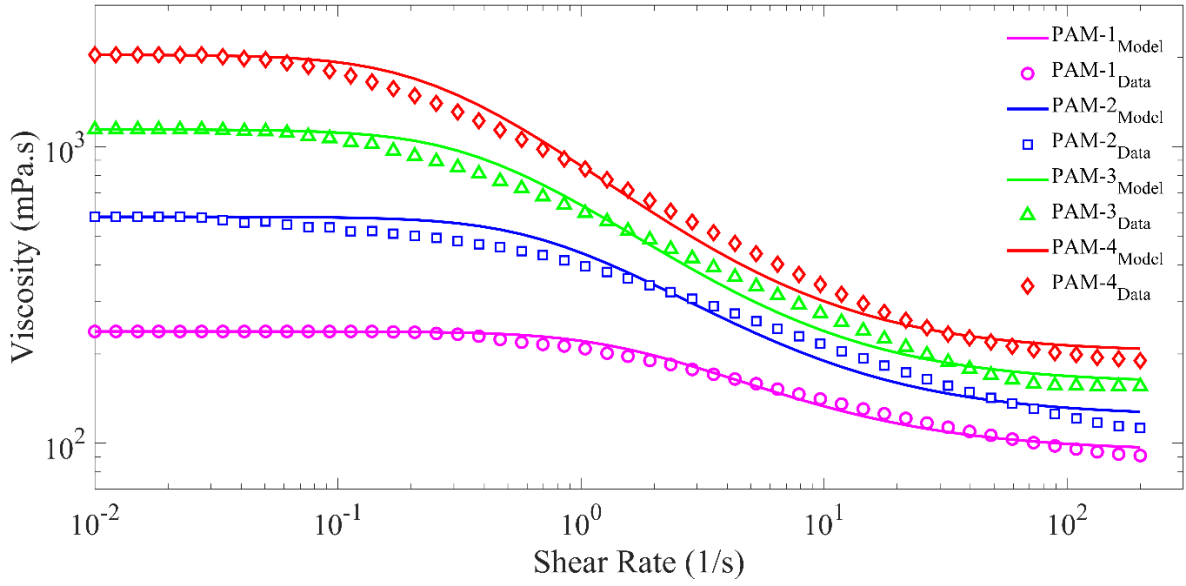

Fig. 3. Viscosity diagrams of four viscoelastic liquids at 25°C. and the optimal response of the Giesekus model.

Table 1. The characteristics of different polymeric solutions at 25°C.

| Sample | $\eta_0^*$ [mPa.s] | $\alpha$ | $\beta$ | $\lambda^*$ [s] | $\Gamma^*$ [mN/m] |
|---|---|---|---|---|---|
| PAM-1 | 238.15 | 0.003 | 0.2832 | 4.2 | 24.87 |
| PAM-2 | 580.20 | 0.001 | 0.1754 | 15.8 | 25.46 |
| PAM-3 | 1142.50 | 0.0016 | 0.1223 | 25.0 | 26.32 |
| PAM-4 | 2041.30 | 0.0023 | 0.0903 | 32.6 | 27.69 |

## 2. Mathematical approach

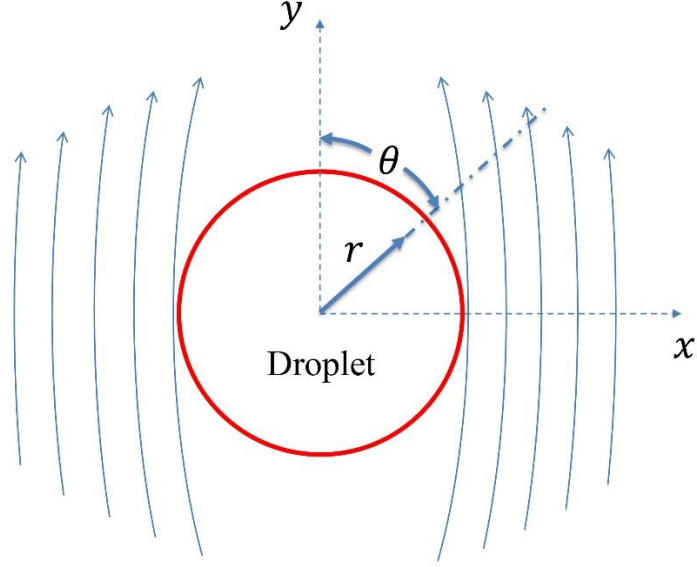

Fig. 4. The geometry of analytical solution

The governing equations of each fluid consists of the continuity and the conservation of momentum that can be expressed as follows:

$$\nabla \cdot \boldsymbol{u}_j^* = 0, \qquad j = i, e, \tag{7}$$

$$\rho_j^* \frac{D\boldsymbol{u}_j^*}{Dt^*} = -\nabla P_j^* + \rho_j^* \boldsymbol{g}_j^* + \nabla \cdot \boldsymbol{\tau}_j^*, \qquad j = i, e. \tag{8}$$

where $\boldsymbol{u}^*$ is velocity vector, $t^*$ is time, $\rho^*$ is the density, $P^*$ is the pressure, $\boldsymbol{g}^*$ is the gravitational acceleration and $\boldsymbol{\tau}^*$ is the stress tensor. Here, the indices "$i$" and "$e$" denote the interior and exterior fluids (i.e., the falling viscoelastic drop and the Newtonian fluid), respectively. The dimensionless parameters employed are defined as follows:

$$De = \frac{\lambda^* U_0^*}{2R^*}, Re = \frac{2\rho_e^* U_\infty^* R^*}{\eta_e^*}, Ca = \frac{\eta_e^* U_\infty^*}{\Gamma^*}, \tag{9}$$

$$\beta = \frac{\eta_i^*{}_{p,0}}{\eta_i^*{}_{t,0}}, \gamma = \frac{\rho_i^*}{\rho_e^*}, k = \frac{\eta_i^*{}_{t,0}}{\eta_e^*}, \eta_{ext} = \frac{\eta_{ext}^*}{\eta_i^*{}_{t,0}},$$

$$\boldsymbol{\tau}_i = \frac{\tau_i^* R^*}{\eta_i^*{}_{t,0} U_0^*}, \boldsymbol{u}_i = \frac{u_i^*}{U_0^*}, \boldsymbol{D}_i = \frac{D_i^* R^*}{U_0^*}, P_i = \frac{P_i^* R^*}{\eta_i^*{}_{t,0} U_0^*},$$

$$\boldsymbol{u}_e = \frac{u_e^*}{U_\infty^*}, \boldsymbol{\tau}_e = \frac{\tau_e^* R^*}{\eta_e^* U_\infty^*}, P_e = \frac{P_e^* R^*}{\eta_e^* U_\infty^*},$$

In Eq. (9), $De$ is the Deborah number (related to the interior medium) and $\lambda^*$ is the corresponding relaxation time, $U_0^*$ is the reference velocity related to the inside of the viscoelastic drop, $R^*$ is the equivalent radius of the drop ($R^* = (\frac{3vol}{4\pi})^{1/3}$, "vol" is volume of falling drop), $Re$ is the Reynolds number, $Ca$ is the capillary number, $U_\infty^*$ is the terminal velocity of the drop, $\Gamma^*$ is the surface tension at the interface between the two fluids, $\beta$ the ratio of the polymeric viscosity $\eta_i^*{}_{p,0}$ to the total viscosity $\eta_i^*{}_{t,0}$ as shear rate tends to zero (also known as the retardation parameter), the density ratio $\gamma$ is defined as the ratio of density of interior $\rho_i^*$ to the exterior fluids $\rho_e^*$, the viscosity ratio $k$ is the ratio of the viscosities of the interior $\eta_i^*{}_{t,0}$ to the exterior fluid $\eta_e^*$, $\eta_{ext}^*$ is extensional viscosity of the interior fluid, and $\boldsymbol{D}_i^*$ is the rate-of-deformation tensor defined as $\boldsymbol{D}_i^* = \frac{1}{2}(\nabla \boldsymbol{u}_i^* + \nabla \boldsymbol{u}_i^{*T})$. Using $k$, a correlation between the reference velocity of the interior medium $U_0^*$ and the terminal velocity $U_\infty^*$ is defined as $U_0^* = \frac{U_\infty^*}{(k+1)}$.

The correlation between stream function and velocity components in the spherical coordinate system is [32-34]:

$$\boldsymbol{u}_j(r,\theta) = \left(\frac{1}{r^2 \sin\theta}\frac{\partial \psi_j}{\partial \theta}, -\frac{1}{r\sin\theta}\frac{\partial \psi_j}{\partial r}\right) \quad j = i, e, \tag{10}$$

### 3.1. Boundary conditions

As a first step, to obtain the stress and velocity distribution, it is assumed that the shape of drop is spherical. Resulting from this assumption, it will be shown that the normal force at the interface won't be balanced so the shape could not be exactly spherical. Following this analysis, a prediction for the shape of the droplets will be obtained to correct this approximation [25]. Assuming a spherical drop, the boundary conditions at the interface of the drop and the external fluid are:

- Two immiscible fluids are selected, so it can be concluded that the radial velocity components on the drop surface are zero:

$$(u_e)_r = 0, (u_i)_r = 0. \tag{11a}$$

- The dimensional tangential velocity and shear stress are equal at the interface due to the cohesion of both fluids. These equalities can be specified as follows (with $k$ denotating the viscosity ratio):

$$(u_e)_\theta = \frac{1}{k+1}(u_i)_\theta. \tag{11b}$$

$$(\tau_e)_{r\theta} = \frac{k}{k+1}(\tau_i)_{r\theta}. \tag{11c}$$

- The normal stress balance at the interface is [35]:

$$\boldsymbol{n}.\left(\boldsymbol{\tau}_e - \frac{k}{k+1}\boldsymbol{\tau}_i\right).\boldsymbol{n} - p_e + p_i = \frac{1}{Ca}\left(\frac{1}{R_1} + \frac{1}{R_2}\right). \tag{11d}$$

- The drop is assumed to fall in an infinite media, while the origin of the coordinate system is fixed at the mass center of the drop. The velocities of the fluid at the drop center and at infinity must be finite:

$$u_i|_{r\to 0} = \text{finite}, \tag{11e}$$

$$u_e|_{r\to\infty} = 1. \tag{11f}$$

Due to the difference between the viscosities of the interior and exterior fluids, different reference velocities are employed for scaling the velocities of each fluid. In Eq. (11d), $\boldsymbol{n}$ is the unit vector normal to the drop surface, and $R_1$ and $R_2$ are the principle radii of the curvature of the surface of the droplet which are unknown parameters and describe the shape of the droplet. In small deformation cases, the following relationship is justified [36]:

$$\frac{1}{R_1}+\frac{1}{R_2}=2-2\zeta-\frac{d}{d\mu}\left((1-\mu^2)\frac{d\zeta}{d\mu}\right), \tag{12}$$

where $\zeta$ is the deformation function and $\mu = \cos(\theta)$. Eventually, the normal stress balance condition in a dimensionless form can be re-written as:

$$\boldsymbol{n}.\left(\boldsymbol{\tau}_e - \frac{k}{k+1}\boldsymbol{\tau}_i\right).\boldsymbol{n} - p_e + p_i = \frac{1}{Ca}\left(2-2\zeta-\frac{d}{d\mu}\left((1-\mu^2)\frac{d\zeta}{d\mu}\right)\right). \tag{13}$$

This correlation will be used later to find the shape of the droplet, i.e. $\zeta$ is the deviation of the droplet shape from spherical.

### 3.2. Perturbation method

In this work, a perturbation method is used to linearize and solve the non-linear governing equations. Using this technique, solutions for stream functions, terminal velocity, pressure and stress tensors can be obtained analytically. Here, a triple perturbation method is used, and the Reynolds, Deborah and capillary numbers are considered as the perturbation parameters. The perturbed forms of the velocity, pressure, deformation tensor, stress tensors and the stream functions can be written as follows:

$$B_j = (B_j)_0 + Re(B_j)_1 + ReCa(B_j)^{(1)} + De(B_j)_{11} + DeCa(B_j)^{(11)},$$
$$B_j = \boldsymbol{u}_j, P_j, \boldsymbol{D}_j, \boldsymbol{\tau}_j, \psi_j \qquad j = i,e. \tag{14}$$

By applying the boundary conditions, the solution of the zeroth order equation is obtained, which is identical with that of Hadamard [20] for Newtonian droplets:

$$(\psi_i)_0 = \left(\frac{r^4}{4}-\frac{r^2}{4}\right)\sin(\theta)^2, \tag{15}$$

$$(\psi_e)_0 = \frac{1}{4}\left(-\frac{(3k+2)r}{k+1}+2r^2+\frac{k}{(k+1)r}\right)\sin(\theta)^2. \tag{16}$$

Here, the normal stress balance boundary condition in Eq. (11d) is expressed as:

$$\delta p = \frac{1}{Ca}\left(\frac{1}{R_1} + \frac{1}{R_2}\right), \tag{17}$$

where, $\delta p$ is the constant ambient pressure jump (not a function of the independent variable $\theta$), so for zeroth order Newtonian term, the drop shape will be spherical. Hence, we have:

$$\delta p = \frac{2}{Ca}. \tag{18}$$

This process is repeated for the first order terms that includes the Reynolds number (further detail can be found in Appendix A). By applying the boundary conditions, the solutions to stream functions with the order of *Re* are obtained:

$$(\psi_i)_1 = -\frac{1}{2}\left(-\frac{(3k+2)r^4}{16(k+1)} + \frac{(3k+2)r^2}{16(k+1)}\right)\sin(\theta)^2 - \frac{1}{2}\left(\frac{(12k^2+23k+10)r^5}{80(k+1)^2} - \frac{(12k^2+23k+10)r^3}{80(k+1)^2}\right)\sin(\theta)^2\cos(\theta), \tag{19}$$

$$(\psi_e)_1 = -\frac{1}{2}\left(\frac{(3k+2)^2 r}{16(k+1)^2} - \frac{3k+2}{8(k+1)}r^2 - \frac{k(3k+2)}{16(k+1)^2 r}\right)\sin(\theta)^2 - \frac{1}{2}\left(\frac{3k+2}{8(k+1)}r^2 - \frac{(3k+2)^2 r}{16(k+1)^2} - \frac{k(3k+2)}{16(k+1)^2 r} + \frac{k(15k^2+22k+8)}{80(k+1)^3} + \frac{k(15k^2+28k+12)}{80(k+1)^3 r^2}\right)\sin(\theta)^2\cos(\theta). \tag{20}$$

The boundary condition presented for the normal stress balance (Eq. (13)) can be presented as:

$$\delta p - 4\alpha_2 Re P_2(\mu) = \frac{1}{Ca}\left(-2\zeta - \frac{d}{d\mu}\left[(1-\mu^2)\frac{d\zeta}{d\mu}\right]\right), \tag{21}$$

where $P_2(\mu) = \frac{(3\mu^2-1)}{2}$ and $\zeta$ is an unknown function which indicates the deviation of the drop shape from a spherical shape. The function $\zeta$ should satisfy the following two conditions:

$$\int_{-1}^{1}\zeta d\mu = 0, \int_{-1}^{1}\zeta\mu d\mu = 0. \tag{22}$$

The above equation guarantees that the center of mass stays at the drop center and the mass conservation is satisfied. In Eq. (21), $\delta p$ and $\alpha_2$ are obtained as:

$$\delta p = \frac{2}{Ca}, \tag{23}$$

$$\alpha_2 = \frac{1}{1920} \frac{(-243k^3 + 20\gamma k - 684k^2 + 20\gamma - 638k - 200)}{(k+1)^3}. \tag{24}$$

Finally, the deformation function for the order of $Re$ ($\zeta_1$) can be obtained as:

$$\zeta_1 = \frac{ReCa}{1920} \frac{(-243k^3 + 20\gamma k - 684k^2 + 20\gamma - 638k - 200)(3\cos(\theta)^2 - 1)}{(k+1)^3}. \tag{25}$$

The stream functions after perturbing around the $ReCa$ parameter are obtained as:

$$(\psi_i)^{(1)} = -\frac{1}{2}\left(\frac{\alpha_2(4-2k)r^4}{5(k+1)} + \frac{(3k-3)r^2}{5(k+1)}\right)\sin(\theta)^2 - \frac{1}{8}\left(\frac{6\alpha_2(5)r^6}{35} + \frac{6\alpha_2(-12)r^4}{35}\right)\sin(\theta)^2(5\cos(\theta)^2 - 1), \tag{26}$$

$$(\psi_e)^{(1)} = -\frac{1}{2}\left(\frac{\alpha_2}{10(k+1)^2}\left[(3k^2 - k + 8)r + \frac{3k^2 - 3k + 6}{r}\right]\right)\sin(\theta)^2 - \frac{1}{8}\left(\frac{\alpha_2}{35(k+1)}\left[\frac{21k+18}{r} - \frac{21k+4}{r^3}\right]\right)\sin(\theta)^2(5\cos(\theta)^2 - 1), \tag{27}$$

Using the normal stress balance (Eq. 13) for the order of $ReCa$, we have:

$$-10\beta_3 ReCaP_3(\mu) = \frac{1}{Ca}\left(2 - 2\zeta - \frac{d}{d\mu}\left[(1 - \mu^2)\frac{d\zeta}{d\mu}\right]\right), \tag{28}$$

where

$$P_3(\mu) = \frac{5\mu^3 - 3\mu}{2}, \quad \beta_3 = \frac{3\alpha_2(-173k - 142)}{700(k+1)}. \tag{29}$$

By applying Eq. (22) and solving Eq. (28), it can be concluded that:

$$\zeta^{(1)} = -\frac{3\alpha_2 ReCa^2(173k + 142)}{700(k+1)} P_3(\mu). \tag{30}$$

The perturbation solution with the order of $De$ can be found:

$$(\psi_i)_{11} = -\frac{1}{2}\left(-\frac{3(\alpha-4)k\beta r^5}{10(k+1)} + \frac{3(\alpha-4)k\beta r^3}{10(k+1)}\right)\sin(\theta)^2\cos(\theta), \tag{31}$$

$$(\psi_e)_{11} = -\frac{1}{2}\left(\frac{3(\alpha-4)k\beta}{10(k+1)^2 r^2} - \frac{3(\alpha-4)k\beta}{10(k+1)^2}\right)\sin(\theta)^2\cos(\theta), \tag{32}$$

Using the normal stress boundary condition and calculating both the normal stress and pressure terms, the shape of the drop can be presented as:

$$\delta p - 4\alpha_3 De P_2(\mu) = \frac{1}{Ca}\left(2 - 2\zeta - \frac{d}{d\mu}\left[(1-\mu^2)\frac{d\zeta}{d\mu}\right]\right), \tag{33}$$

therefore:

$$\delta p = \frac{2}{Ca}, \tag{34}$$

$$\alpha_3 = \frac{1}{60}\frac{\beta(54k^2\alpha + 97k\alpha - 65)}{(k+1)^2}. \tag{35}$$

The deformation function for the order of $De$ can be presented as:

$$\zeta_{11} = -\frac{DeCa}{60}\frac{\beta(54k^2\alpha + 97k\alpha - 65)(3\cos(\theta)^2 - 1)}{(k+1)^2}. \tag{36}$$

The Capillary-Deborah term is obtained in a manner like the Reynolds-Capillary term, but for brevity they are presented in Appendix A (i.e. Eqs. (A20) and (A21)). Using the normal stress balance boundary condition, the final drop shape is calculated as following:

$$R(\mu) = 1 - \alpha_2 ReCa P_2(\mu) - \beta_3 ReCa^2 P_3(\mu) - \alpha_3 DeCa P_2(\mu) - \beta_3 DeCa^2 P_3(\mu). \tag{37}$$

For the internal and external fluids, the pressure functions can be calculated through substituting the obtained velocity functions in the momentum equation (not presented for brevity here but presented in Appendix A). As a next step, the drag coefficient and the terminal velocity are calculated. In order to calculate the terminal velocity, the overall drag value and the apparent drop weight need to be balanced. The overall force applied on the drop is $F^* = F_A^* - F_D^*$, where $F_A^*$ is the apparent drop weight (i.e. the difference between the weight and buoyancy forces) and $F_D^*$ is the drag force. The apparent drop weight ($F_A^*$) is calculated as:

$$F_A^* = \frac{4\pi \rho_e^* g^* R^*(1-\gamma)}{3 U_\infty^{*\,2}}. \tag{38}$$

The drag force can be found as:

$$F_D^* = 2\pi R^{*2} \int_{-1}^{1}[\mu((\tau_e^*)_{rr} - \mu P_e^* - (1-\mu^2)^{1/2}(\tau_e^*)_{r\theta})]|_{r=1} d\mu. \tag{39}$$

By calculating the above integral, we have:

$$F_D^* = \eta_e^* U_\infty^* R^*\left(2\pi \frac{3k+2}{k+1} + Re\frac{\pi}{8}\left(\frac{3k+2}{k+1}\right)^2 + \frac{2\pi\alpha_2 ReCa}{10(k+1)^2}(3k^2 + 11k - 4) + \frac{2\pi\alpha_3 DeCa}{10(k+1)^2}(3k^2 + 11k - 4)\right). \tag{40}$$

The drag coefficient is calculated using $C_D = \frac{F_D^*}{\frac{1}{2}\rho_e^* U_\infty^{*\,2} A^*}$, where $A^* = \pi R^{*2}$ is the approximate area of projection surface of the drop. The drag coefficient is calculated as follows:

$$C_D = \frac{4}{\pi Re}\left(2\pi \frac{3k+2}{k+1} + Re\frac{\pi}{8}\left(\frac{3k+2}{k+1}\right)^2 + \frac{2\pi\alpha_2 ReCa}{10(k+1)^2}(3k^2 + 11k - 4) + \frac{2\pi\alpha_3 DeCa}{10(k+1)^2}(3k^2 + 11k - 4)\right). \tag{41}$$

In the limiting case of $k \to \infty$, $Ca = 0$ and $De = 0$, the above equation is simplified to the well-known solution of Oseen [37] for falling an immersed sphere through a viscous liquid in the inertial regime:

$$C_D = \frac{24}{Re}\left(1 + \frac{3}{16}Re + \cdots\right), \tag{42}$$

which shows the consistency of the presented solution with the available literature results.

In the steady state condition, the resultant overall force exerted on to the drop is zero ($\sum F^* = 0$). In this condition, the drag force is balanced with the apparent drop weight. Therefore, it is possible to determine the drop terminal velocity by balancing these two forces (Eq. (38) and (40)) as following:

$$U_\infty^* = \frac{4\pi}{3} \frac{\rho_e^* g^* R^{*2}(\gamma-1)}{\eta_e^*\left(2\pi\frac{3k+2}{k+1}+Re\frac{\pi}{8}\left(\frac{3k+2}{k+1}\right)^2+\frac{2\pi\alpha_2 ReCa}{10(k+1)^2}(3k^2+11k-4)+\frac{2\pi\alpha_3 DeCa}{10(k+1)^2}(3k^2+11k-4)\right)}. \tag{43}$$

Applying *Re*=0 and *De*=0, the above solution is simplified to the well-known solution of Hadamard [20] and Rybczynski [21] for the creeping motion limit of a falling Newtonian drop into another immiscible Newtonian fluid:

$$U_\infty^* = \frac{2(k+1)}{3(3k+2)} \frac{\rho_e^* g^* R^{*2}(\gamma-1)}{\eta_e^*}. \tag{44}$$

Which is again consistent with the results available in the literature.

## 4. Results and Discussion

### 4.1. Drop deformation

The analytical results obtained for the shape of droplets are validated against the experiments of falling viscoelastic drops through Newtonian media in the low inertia regime.

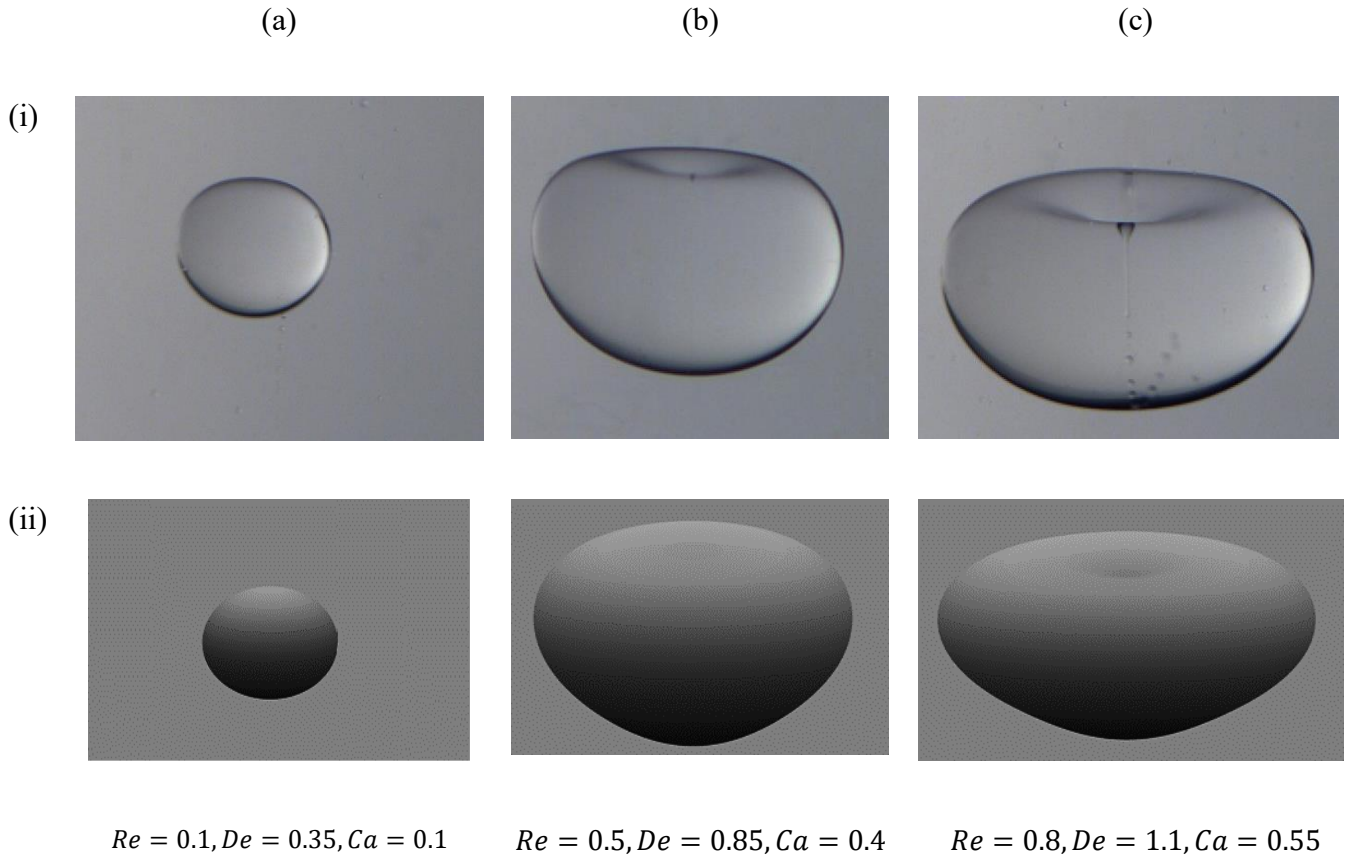

$Re = 0.1, De = 0.35, Ca = 0.1$  $\quad Re = 0.5, De = 0.85, Ca = 0.4 \quad$  $Re = 0.8, De = 1.1, Ca = 0.55$

Fig. 5. Comparison between steady shapes of viscoelastic drops (PAM-2), obtained from (i) the experimental observations and (ii) the analytical solution. The drop volumes are: (a) 0.4 ml, (b) 1 ml and (c) 1.4 ml.

Fig. 5 shows the comparison made between experimental results and analytical solutions (presented in 3D using the axis-symmetry assumption for comparison purposes) for the steady shape of drops at different Reynolds, Deborah and capillary numbers. Here, the PAM-2 fluid is used in the experimental section ($k = 0.97, \alpha = 0.001, \beta = 0.1754$ and $\gamma = 1.3$). The results include a range of small to large drops with the volume of 0.4, 1 and 1.4 ml. The presented results suggest that the analytical solution, can capture the important physics involved during the shape transformation of viscoelastic droplets in good agreement with the experiments. Note that in Fig. 5(c)(i) there appears to be an internal thread that is breaking into droplets. Those drops are all coming back up in one direction suggesting not quite symmetric flow internally. In this experiment, by increasing the size of the droplet the terminal velocity of the droplet increases (Equations 43-44) leading to an increase in both the elastic and inertia forces. This issue suggests that a different experimental protocol than a change on the droplet size would be required to separate out the influence of each inertia and viscoelastic forces on the drop deformation problem.

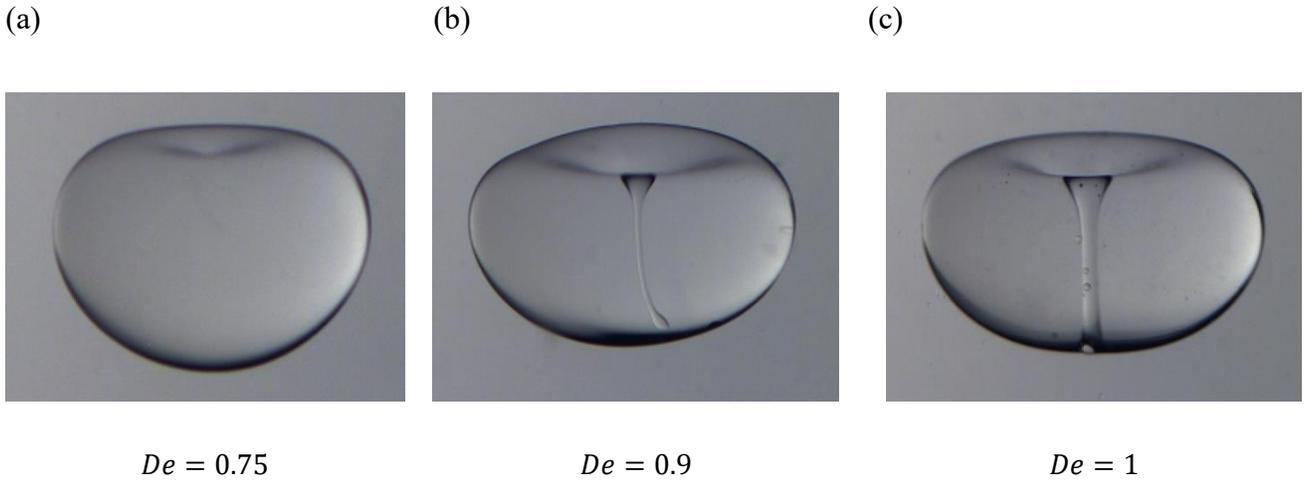

(a) $De = 0.75$  (b) $De = 0.9$  (c) $De = 1$

Fig. 6. Steady-state shapes of (a) PAM-1, (b) PAM-3 and (c) PAM-4 drops falling through 600 cP oil ($Re = 0.6\ Ca = 0.5, V^* = 1.2$ ml).

In Fig. 6 three elastic fluids with different relaxation times are used to generate droplets with similar sizes. This approach allows us to obtain different values of Deborah number while the Reynolds and Capillary numbers are "fixed" at constant values (i.e. the terminal velocity of droplets are equal, but different values of relaxation time of these fluids lead to different elastic forces). As shown here, by increasing the viscoelastic properties of the fluid at a fixed Reynolds number, a dimple at the rear end of the droplet starts to grow which looks like an entrainment instability ($C_f$ interface "cusps" Fig. 6c) [38]. The same qualitative results were also reported using Oldroyd-B model previously by [39]. According to our best knowledge, the reason behind this transformation was not previously fully explained and it was simply attributed to the non-linear properties of elastic and inertial forces. Here we use the analytical solution obtained in the previous sections to investigate and study the effect of different flow types on this transformation in more details. A helpful parameter to study the effect of different types of flow on such complex problems would be the flow type parameter $\xi$ defined as [5-7]:

$$\xi = \frac{\|D_j\| - \|\Omega_j\|}{\|D_j\| + \|\Omega_j\|}, \quad j = i, e, \tag{45}$$

where $\|D_j\|$ and $\|\Omega_j\|$ are the magnitude of the rate of deformation and vorticity tensors:

$$D_j = \frac{1}{2}(\nabla u_j + \nabla u_j^T), \quad \Omega_j = \frac{1}{2}(\nabla u_j - \nabla u_j^T) \quad j = i, e. \tag{46a}$$

$$\|D_j\| = \left(\frac{D_j : D_j}{2}\right)^{\frac{1}{2}}, \quad \|\Omega_j\| = \left(\frac{\Omega_j : \Omega_j}{2}\right)^{\frac{1}{2}}, \quad j = i, e, \tag{46b}$$

The $\xi$ parameter may vary within the range $[-1,1]$ in which $\xi$ =-1 characterizes a solid-like rotational flow, $\xi$ =1 a pure extensional flow and $\xi$ =0 a simple shear flow.

As shown in Figure 7(a and c) at low values of inertia and elastic forces, two stagnation points are presented at the top and rear end of the droplet in both fluids which are corresponding to *compressional* and *extensional* regimes respectively. Presence of such stagnation points have been previously reported in flow past solid spheres [40] and cylinders [41]. Throughout this paper we refer to them as the "focal" stagnation points. In Figure 7 (b and d) it is shown that by increasing the Reynolds and Deborah numbers these elongational dominated regimes spread and eventually divides into cylindrical symmetry (a point and a circular line), in 3D presentation, in the exterior media while within the droplet phase, there still exists one main strand of elongational flow. Note that the newly developed stagnation points on the droplet surface in 2D presentation are, in fact, "circular stagnation lines" in 3D. A similar trend of splitting elongational regime was also recently observed and reported by Davoodi et al. [5], at the free and pinned stagnation point of viscoelastic flows in cross-slot geometries. In the current problem, an increase in interfacial tension can once again reunite this newly developed stagnation line to the focal stagnation point. Note that these strands are located at a position where, the shear rate is almost zero (see figure 8), and due to a non-zero value of the streamwise gradient of velocity, a "purely" elongational flow is observed.

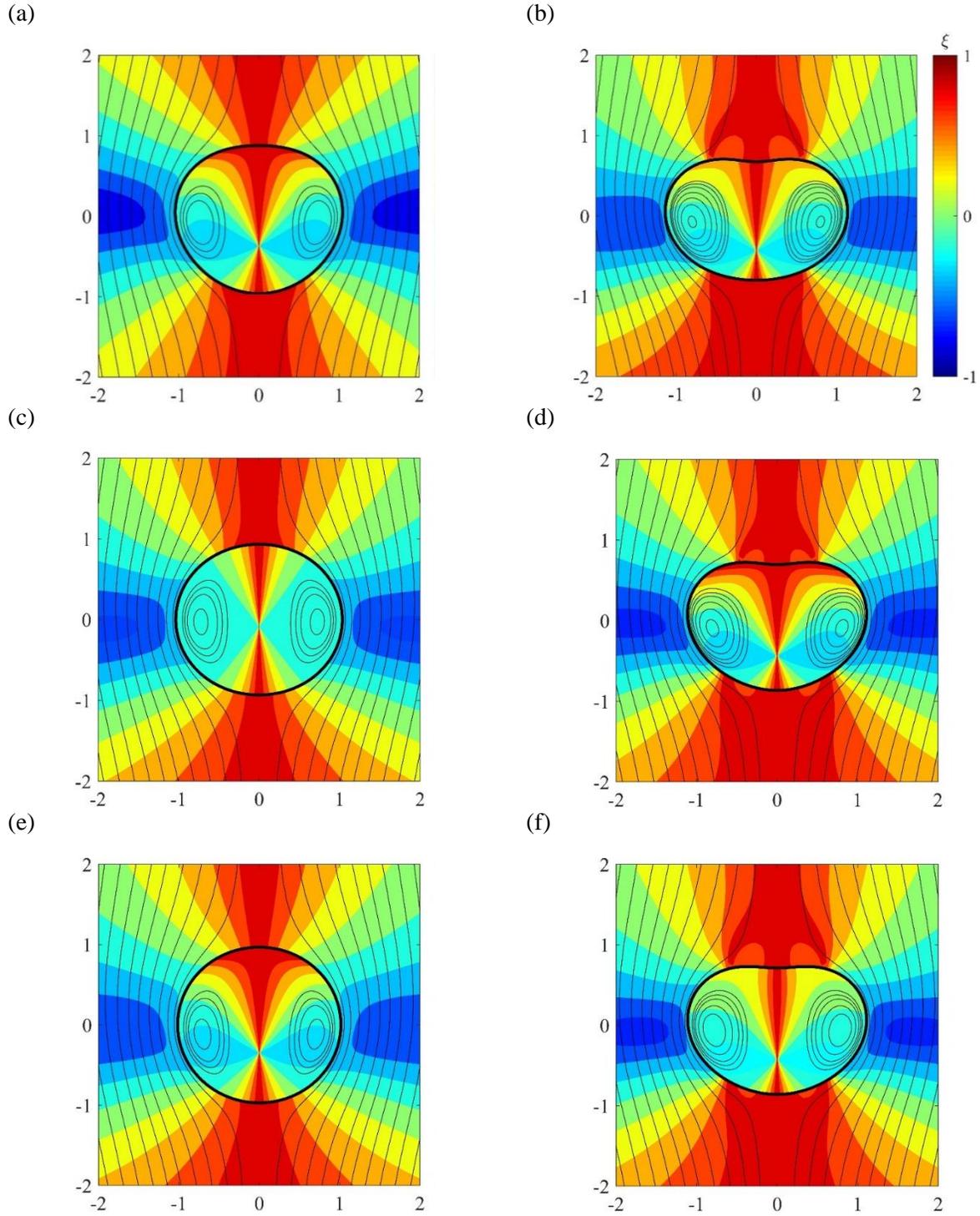

Fig. 7. Stream line and flow-type parameter ($\xi$) contours with superimposed streamlines for different Reynolds (a) $Re=0$, (b) $Re=0.9$ at fixed $De=0.6$, $Ca=0.5$ and different Deborah (c) $De=0$, (d) $De=0.9$ at fixed $Re=0.6$, $Ca=0.5$ and different capillary (e) $Ca=0.1$, and (f) $Ca=0.7$ at fixed $Re=0.6$, $De=0.5$ numbers with $k=1$, $\alpha=0.2$, $\beta=0.5$, $\gamma=1.3$.

Interestingly, these stagnation points/lines appear at local extrema locations of the droplet interface as was previously observed by Davoodi et al. [5] in cross-slot geometries. At the focal stagnation point, located at the rear end of the droplet where a dimple exists, there is a competition between the two normal forces applied from the interior and exterior media which are pulling the surface of the droplet in opposite directions. In such simple steady state elongational flows, the magnitude of these normal forces may be calculated by multiplying the strain rate with the elongational viscosity. The elongational viscosity $\eta_{ext}$ in the uni-axial flows, such as the ones observed at the focal stagnation point, can be defined as:

$$(\eta_{ext})_j = \left(\frac{(\tau_{rr})_j - (\tau_{\theta\theta})_j}{\nabla u_j(1,1)}\right)\Big|_{r=1,\cos(\theta)=0}, \quad j = i, e, \qquad (47)$$

One can show that, unlike the shear thinning properties of viscosity in shear flows for viscoelastic fluids, the elongational viscosity is a shear thickening function of the applied strain rate (more details related to elongational viscosity for the non-linear and linearized Giesekus model can be found in Appendix B).

In Figure 9 the variation of the elongation viscosity as a function of *De* at the central stagnation point is presented. The strain dependency of the *perturbed* Giesekus model predicts a strain-hardening behaviour (the viscosity increases with increasing strain rate) for low strain rates, but for higher strain rates disaster appears; the viscosity diverges $\lambda\dot{\epsilon} = 0.5$ and at higher strain rates, the model predicts a negative value for the steady-state elongational viscosity. Such behaviours have been previously reported for the Oldroyd-B model and not been predicted for the full non-linear Giesekus model. In Appendix B we show that is indeed *a result of the linearization* of the Giesekus model, so care should be taken once such linearization perturbation techniques are used in elongational dominated flows.

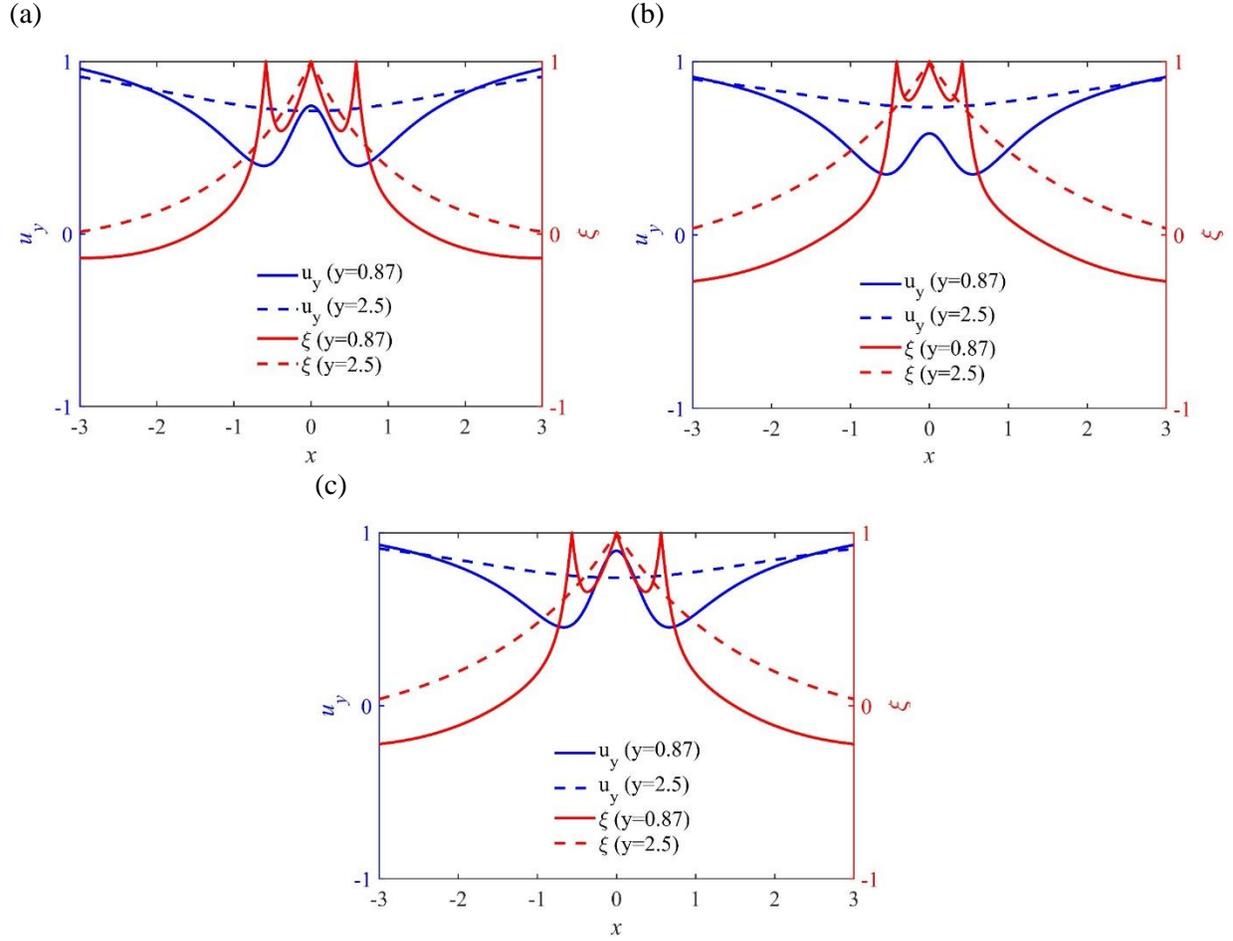

Fig. 8. Comparison of non-dimensional velocity distribution and the flow type parameter along the horizontal line at y=0.87 and y=2.5 for different Reynolds, Deborah and capillary numbers (a) *Re=0.9, De=0.6, Ca=0.5*, (b) *De=0.9, Re=0.6, Ca=0.5* and (c) *Ca=0.7, Re=0.6, De=0.5* with *k*=1, *α*=0.2, *β*=0.5, *γ*=1.3.

As mentioned earlier, further information about elongational viscosity for linearized and full non-linear Giesekus model is presented in appendix B. Knowing that $(\eta_{ext})_{Newtonian} = 3$ [1, 42] and from the results presented in Figure 9, one can realize for any non-zero value of Deborah number the elongational viscosity of interior fluid, i.e. the viscoelastic fluid, is always greater than the Newtonian exterior media i.e. $(\eta_{ext})_e < (\eta_{ext})_i$ (note whilst the Trouton ratio is 3 for a Newtonian fluid, polymeric materials finds a bigger value). Now, simply it can be concluded that on the focal stagnation points with a fixed strain rate, the applied interior normal stress is higher than the exterior stress, i.e. $(\tau_{rr})_e < (\tau_{rr})_i$, therefore the interior fluid is able to pull the interface of the droplet inward at the rear end of the droplet (due to extensional flow) and also to push the interface of the droplets outward at the front of the droplets (due to compressional flow). In other words, the normal forces at the focal stagnation points squeeze the droplet and consequently droplet finds a dimple at the rear end. At this stage once the Deborah or Reynolds number is increased further, a circular line of stagnation point starts to grow, centered around the focal stagnation points, on the exterior media. On the rear end this newly developed circular elongational line will pull the droplet interface outward resulting to the enhance of a dimple. Note that, unlike the rear end, on the front side of the droplet, the exterior media is trying to push the droplet interface inward and this will create a "cusp" like front on the droplet interface.

To investigate the deviation of the drop from spherical shape in a more quantitative manner, an asymmetry parameter $AP$ is defined as follows:

$$AP = \frac{\int_0^{2\pi} \zeta^2 d\theta}{2\pi}, \tag{48}$$

The variation of $AP$ as a function of Deborah, Reynolds and Capilary numbers are presented in Figure 10. As shown, once the interfacial tension is much larger than viscous forces, i.e. $Ca \approx 0$, the droplet is spherical, the $AP$ always goes to zero. Although both inertia and elastic forces can separately increase the deformation of droplet from spherical shape, the results shown in Figure 10 suggest that there is a finite range that these two forces can have opposite effects on the total deformation of the droplet leading to the presence of minimum extremum in these graphs.

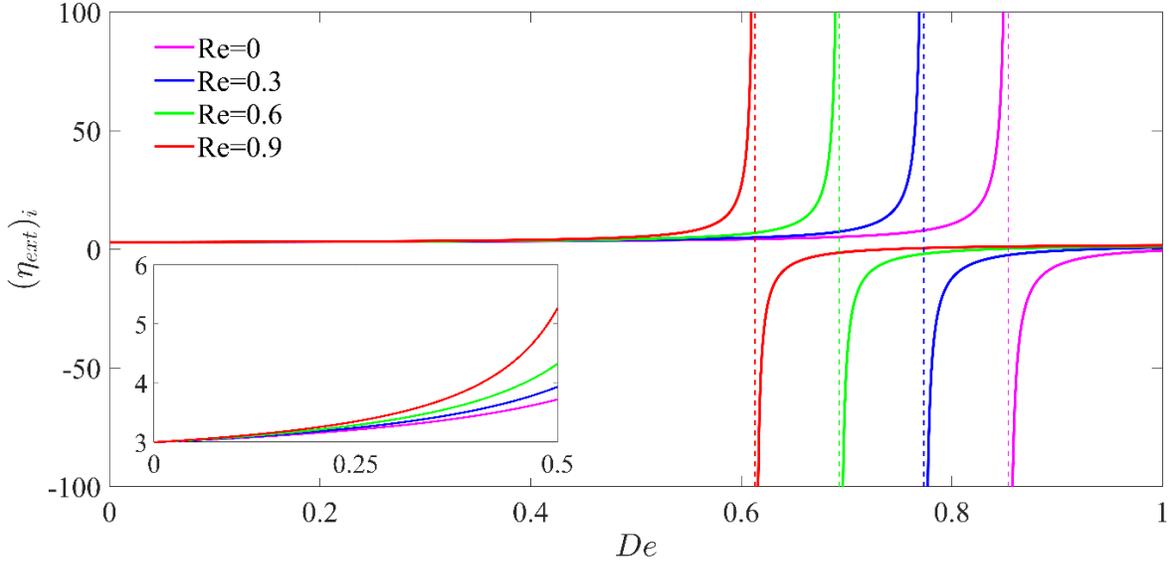

Fig. 9. The variation of $(\eta_{ext})_i$ versus the Deborah number for different Reynolds number at $Ca$=0.5, $k$=1, $α$=0.2, $β$=0.5, $γ$=1.3.

### 4.1.1. Terminal velocity analysis

Setting $Re = 0$ in our formulation will allow us to compare our results with the analytical and experimental results of Sostarecz and Belmonte [25] related to the motion of falling viscoelastic drops through Newtonian media in the creeping regime. In this research [24] a polymeric solution of 0.16% xanthan gum by weight in 80:20 glycerol/water was used as the interior viscoelastic fluid and polydimethylsiloxane oil as the exterior viscous fluid. The properties of interior and exterior fluid in their work are $\rho_i^* = 1.27 \text{ g.cm}^{-3}, (\eta_i^*)_0 = 50 \text{ Pa.s}, \rho_e^* = 0.98 \text{ g.cm}^{-3}, \eta_e^* = 0.98 \text{ Pa.s}, \lambda^* = 40 \text{ s}$, and as a result $k = 50, \alpha = 0.32, \gamma = 1.3$. Here, the mobility factor is estimated by curve fitting on the viscometric data reported by Sostarecz and Belmonte [25]. A comparison between the terminal velocities (from our study and Sostarecz and Belmonte [25]) for different equivalent drop radii is presented in Table 2. According to the Table, the result of our analytical study presents a better fit with their experimental data [25].

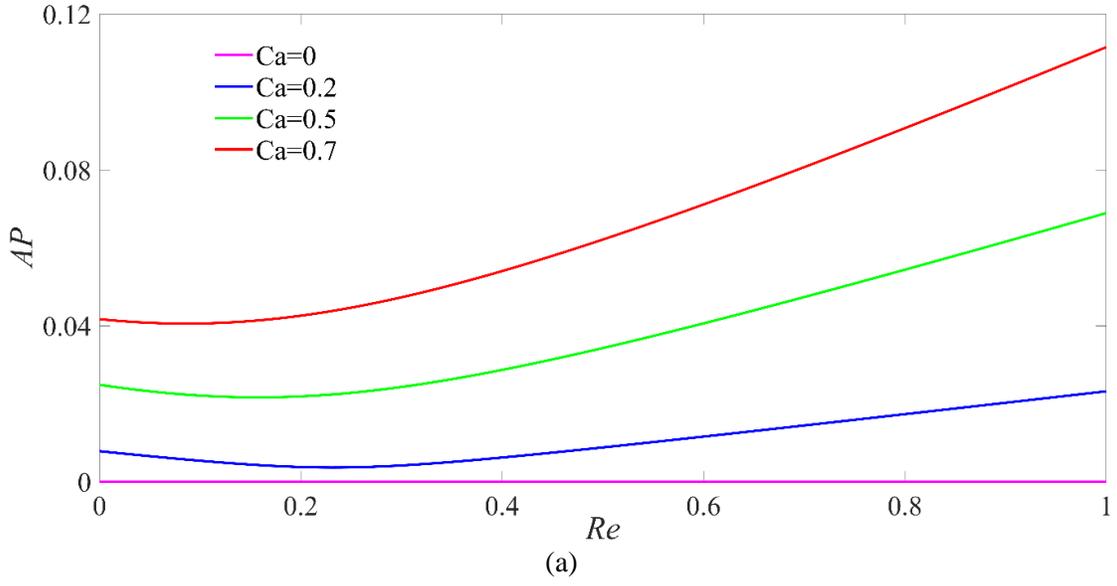

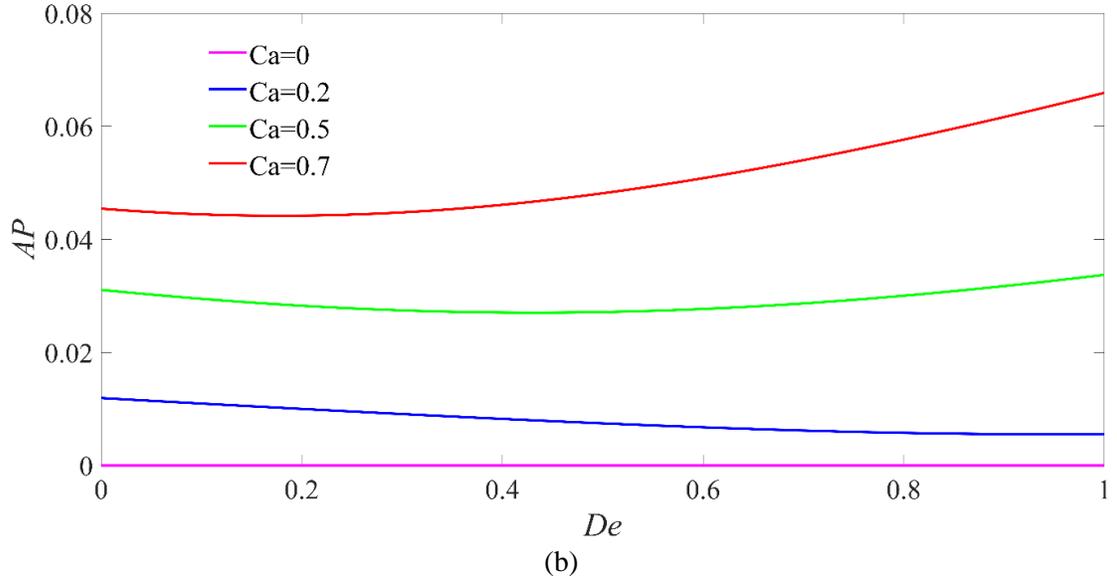

Fig. 10. Variation of the asymmetry parameter (a) versus the Reynolds number at *De*=0.7, *k*=10, *α*=0.2, *β*=0.5, *γ*=1.3 and (b) versus the Deborah number at *Re*=0.5, *k*=40, *α*=0.3, *β*=0.7, *γ*=1.3 for different values of *Ca* number.

The larger difference between the analytical solution of Sostarecz and Belmonte [25] and their experimental data may be attributed to the simplification using the Payne–Pell [43] theorem in their work. Note that in our work, Eq. (39) was used for the calculation of the drag force. Following this validation in Figure 11a, the variation of the terminal velocity versus the elasto-capillary number for different mobility factors ($\alpha$) is presented. The elasto-capillarity is the ability of the capillary force to deform an elastic material ($Ec = \frac{De}{Ca} = \frac{\lambda^* \Gamma^*}{(k+1)\eta_e^* R^*}$). This dimensionless parameter is defined solely based on the fluid properties [44] and it is independent of the kinematic parameters such as the velocity and the shear rate. It is observed that by increasing the mobility factor ($\alpha$), the drop terminal velocity increases.

**Table. 2.** A comparison between results of present study and work of Sostarecz and Belmonte [25] for terminal velocity of falling viscoelastic drop in Newtonian media ($\delta_1$ is the percentage of error between the experimental results of Ref. [25] and their analytical solution, and $\delta_2$ is the percentage of error between the experimental results of Ref. [25] and the analytical solution obtained in the current study).

| $R^*$ [cm] | $U_\infty^*$[cm/s] (Experimental Ref. [25]) | $U_\infty^*$[cm/s] (Analytical Ref. [25]) | $U_\infty^*$[cm/s] (Analytical present study) | $\delta_1$[%] | $\delta_2$[%] |
|---|---|---|---|---|---|
| 0.067 | 0.024 | 0.02603 | 0.01997 | 8.46 | 16.79 |
| 0.12 | 0.085 | 0.08873 | 0.06806 | 4.39 | 19.93 |
| 0.15 | 0.102 | 0.1386 | 0.1064 | 35.88 | 4.31 |
| 0.16 | 0.1 | 0.1577 | 0.1210 | 57.70 | 21.00 |
| 0.173 | 0.125 | 0.1834 | 0.1427 | 46.72 | 14.16 |
| 0.19 | 0.15 | 0.2224 | 0.1706 | 48.27 | 13.73 |
| 0.225 | 0.2 | 0.312 | 0.2393 | 56.00 | 19.65 |
| 0.24 | 0.265 | 0.3549 | 0.2723 | 33.92 | 2.75 |
| 0.285 | 0.31 | 0.5005 | 0.3839 | 61.45 | 23.84 |
| 0.3 | 0.353 | 0.5546 | 0.4254 | 57.11 | 20.51 |

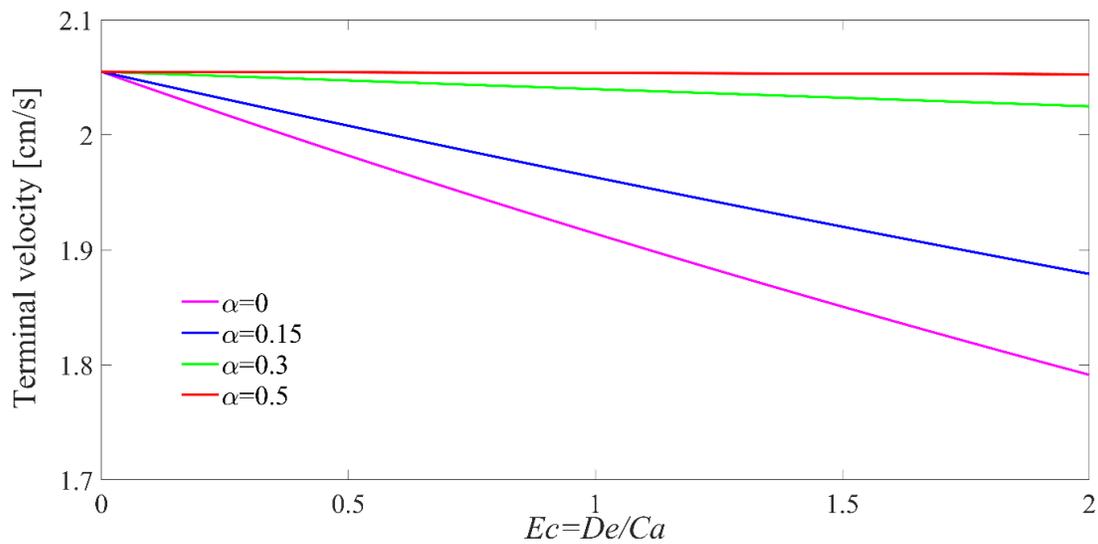

(a)

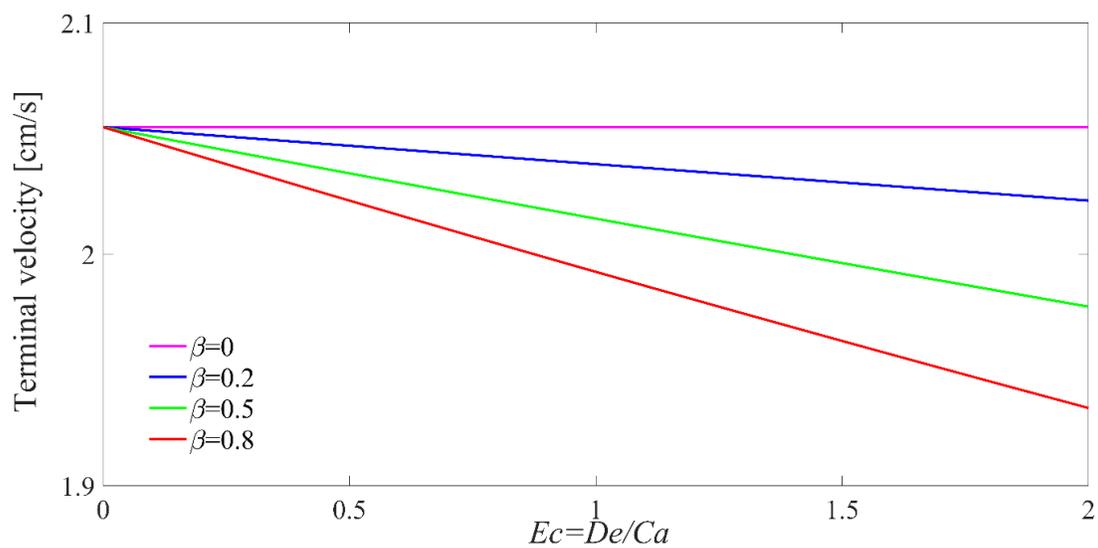

(b)

Fig. 11. Terminal velocity versus the elasto-capillary number for (a) different mobility factors ($\alpha$) at $Re = 0.8, \beta = 0.5, k = 1, \gamma = 1.3, R^* = 0.5$ cm, $\rho_e^* = 0.9$ g. cm$^{-3}$, $\eta_e^* = 0.6$ Pa. s  and  (b) different retardation ratios at $Re = 0.9, \alpha = 0.2, k = 1, \gamma = 1.3, R^* = 0.5$ cm, $\rho_e^* = 0.9$ g. cm$^{-3}$, $\eta_e^* = 0.6$ Pa. s.

This effect arises from the fact that an increase in $\alpha$ will lead to a decrease of the effective shear viscosity of the interior fluid and this will consequently lead to the reduction of the drag force on the surface of the droplet. According to the figure, at small mobility factors, the terminal velocity decreases when increasing the elasto-capillary number. At large mobility factors, such as $\alpha = 0.5$, the terminal velocity is approximately independent of the elasto-capillary number and the elastic and surface tension forces are balanced. The diagram of the terminal velocity versus the elasto-capillary number for different values of $\beta$ is shown in Fig. 11b, where the terminal velocity decreases by increasing $\beta$. This feature can be attributed to the drop deformation and increasing the drop surface projection area that increases the frictional and pressure drag forces.

### 4.1.2. Drag coefficients analysis

For a more quantitative comparison, the drag coefficients obtained from experimental observations and analytical solutions are compared with each other. The resultant force of a falling drop, $F^*$, ($F^* = F_A^* - F_D^*$) where $F_A^*$ is the apparent weight of drop (the difference between the weight and Buoyancy) and $F_D^*$ is drag force which can be expressed as follows:

$$F_A^* = (\rho_i^* - \rho_e^*)\mathbf{g}^* V^*, \tag{49}$$

$$F_D^* = \frac{1}{2}\rho_e^* U_\infty^{*\,2} A^* C_D. \tag{50}$$

where $A^*$, $V^*$ are the projection area and volume of droplet. In steady state situation, the force exerted on the drop is zero ($F^* = 0$). By balancing the apparent weight and drag force from Eqs. (49) and (50), the drag coefficient calculated in our experiments is as following:

$$(C_D)_{Ex} = \frac{8}{3}\frac{\mathbf{g}^* R^*(\gamma-1)}{U_\infty^{*\,2}}. \tag{51}$$

By calculating the terminal velocity from the image processing of captured figures and using Eq. (51), the experimental drag coefficient can be calculated. Fig. 12 depicts the variation of the drag coefficient versus the Reynolds number for four types of polymeric drops which are used in our study. The results are obtained at low Reynolds inertia regime and the analytical drag coefficient are calculated using Eq. (41) based on the properties which are reported in Table 1 for PAM-1 to PAM-4. A good agreement between the analytical solution and experimental data is observed, the small deviation between the results could be attributed to the error of perturbation method (as an approximation method) and empirical uncertainties.

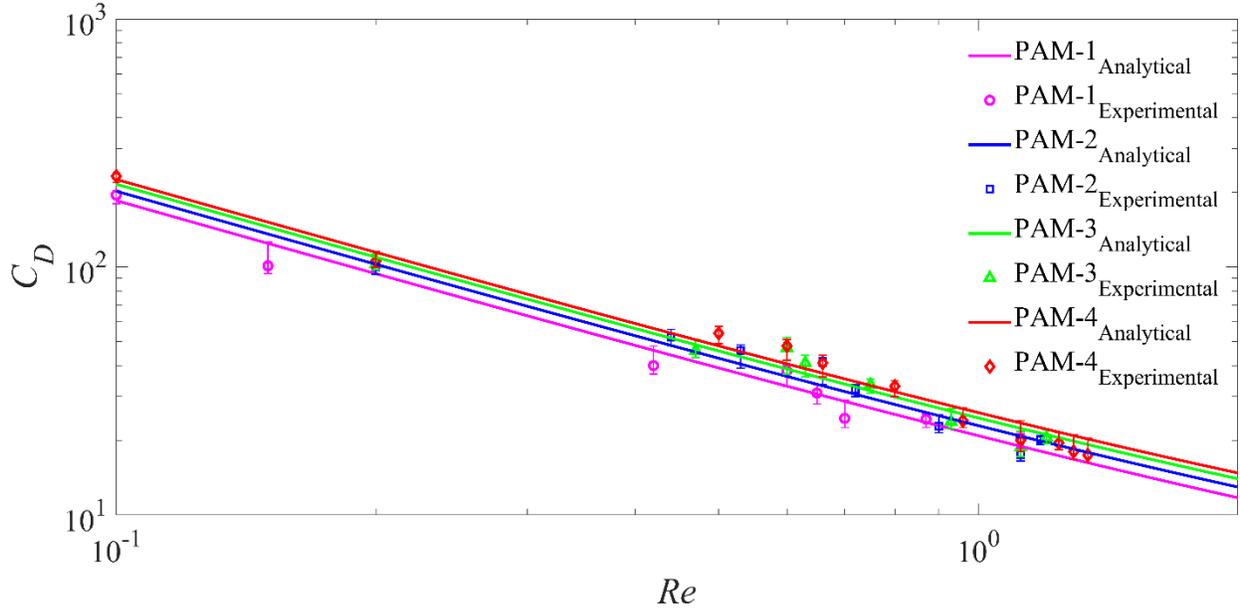

Fig. 12. The diagrams of drag coefficient versus the Reynolds number for four viscoelastic liquids at *De*=0.6, *Ca*=0.5, *γ*=1.3.

Figure 13 shows the drag coefficient diagram versus elasto-capillary number for different Reynolds numbers. It is shown that the drag coefficient increases by increasing both the Deborah and capillary numbers. As shown, the drag coefficient is increased as drop is losing its spherical shape. This effect can be attributed to increase of the drop surface area at a constant volume, which loosely speaking, causes more friction. The pressure drag force is also enhanced by the drop deformation, due to increasing the projected surface area.

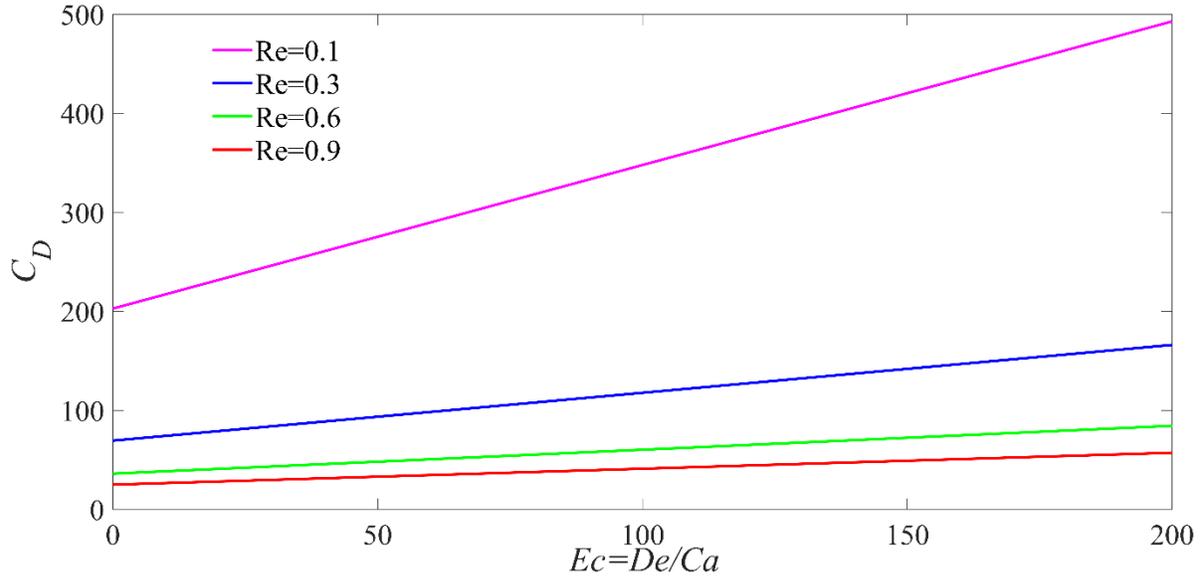

Fig. 13. The drag coefficient versus the elasto-capillary number at the interface for different Reynolds numbers at $\alpha = 0.2, \beta = 0.5, k = 1, \gamma = 1.3$.

Figure 14 illustrates the effects of the Reynolds number on the drag coefficient of falling viscoelastic drops at $De = 0.8, Ca = 0.5, \alpha = 0.2, k = 1, \beta = 0.5, \gamma = 1.3$. The Drag force applied on the surface of the droplet consists of pressure and shear stress contributions. As expected, the drag coefficient decreases by increasing the Reynolds number and the friction contribution is more than the pressure contribution in this regime. In order to better understand the effects of different parameters on the drag coefficient, the contribution of perturbation parameter on the drag coefficient is presented in Table 3. According to the Table, most of the contribution in the drag force arises from the zeroth order term. The total contribution of the inertial term on the drag coefficient (i.e., the terms with order of *Re* and *ReCa*) is around 17%. It is worth mentioning that most of the previous analytical solutions on the dynamics of falling viscoelastic drops through Newtonian media have been restricted to the creeping motion regime. Thus, based on our results, the drag coefficient errors in the previously reported works would be considerable at the low Reynolds regimes, due to neglect of the inertial terms.

**Table 3.** The contribution of each part of the drag force terms at $De = 0.8, Ca = 0.5, \alpha = 0.2, k = 1, \beta = 0.5, \gamma = 1.3$ (Eq. (41)).

| Order of Perturbation Terms | Percentage (%) |
|---|---|
| Zeroth | 70.3 |
| $Re$ | 13.1 |
| $ReCa$ | 3.9 |
| $DeCa$ | 12.7 |

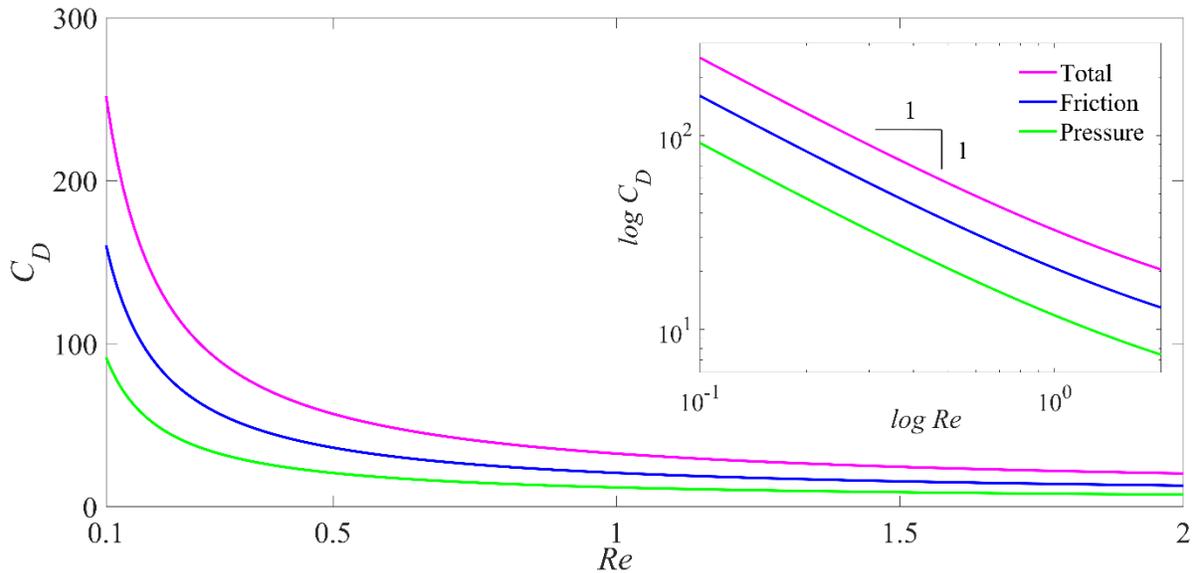

Fig. 14. The drag coefficient versus the Reynolds number at $De = 0.8, Ca = 0.5, \alpha = 0.2, \beta = 0.5, k = 1$ and $\gamma = 1.3$ (Eq. (41)).

## 5. Conclusions

In the paper, an analytical solution based on a triple perturbation method is presented for an immiscible flow in which viscoelastic drops fall into Newtonian media at low Reynolds numbers. Using the Giesekus model, a solution is derived for the shape of viscoelastic drops falling into quiescent media. These results are validated against our experimental observations using polymeric solutions for the drop phase.

Our results suggest that as Deborah number or Reynolds number is increased a circular stagnation line starts to appear, centered around the focal stagnation. A similar type of splitting of elongational dominated flows in the cross-slot geometries for single phase flows has been also previously reported by Davoodi et [5]. As the size of the droplet is increased the terminal velocity of the droplet and so the strain rate at the focal stagnation points are increased and due to the shear thickening properties of the elongational viscosity in viscoelastic fluids, the droplet finds a dimple at the rear end and the pointy shape at the front. The presence of this newly unveiled circular stagnation line at the rear end of the droplet can pull the droplet interface outward resulting in the enhancement of dimple appearance. It has also been shown that increasing the mobility factor weakens the shear-thickening behaviors of the elongational viscosity and decreases the radial normal stress especially within the elongational dominated regions. Therefore, the dimple size decreases.

Also, due to singularity issues appearing in elongational dominated flows, use of non-linear models such as Giesekus model has been recommended over Oldroyd-B constitutive equation. In this study, we have shown once the non-linear Giesekus constitutive equations is linearized using a perturbation method, similar singularity issues might arise, and special care should be taken to avoid cases with $\dot{\epsilon} \to 0.5/\lambda$ and the perturbation parameters should be limited to problems with $Wi < 0.5$.

**Acknowledgment**

**Appendix A**

The zeroth and first order terms of the stress tensor for interior and external fluids are:

$$(\tau_j)_0 = 2(D_j)_0 \quad j = i, e, \tag{A1}$$

$$(\tau_j)_1 = 2(D_j)_1 \quad j = i, e. \tag{A2}$$

After collecting the terms at the zeroth order of the Reynolds number, we have:

$$E^4(\psi_j)_0 = 0 \quad j = i, e, \tag{A3}$$

where $E^4 = E^2 E^2$ and $E^2$ is defined as follows:

$$E^2 = \frac{\partial^2}{\partial r^2} + \frac{\sin\theta}{r^2}\frac{\partial}{\partial \theta}\left(\frac{1}{\sin\theta}\frac{\partial}{\partial \theta}\right). \tag{A4}$$

Happel and Brenner[45] showed that the solution to Eq. (A3) has the following form:

$$\psi_j(r,\theta) = \sum_{n=1}^{\infty}(A_n r^{-n} + B_n r^{2-n} + C_n r^{n+1} + D_n r^{n+3})Q_n(\mu) \quad j = i, e, \tag{A5}$$

where $Q_n(\mu)$ is the Gegenbauer polynomial, and it correlates with the Legendre polynomial as:

$$Q_n(\mu) = \int_{-1}^{\mu} P_n(s)\, ds. \tag{A6}$$

The equations of the first order $Re$ stream functions are written as:

$$E^4(\psi_i)_1 = 0, \tag{A7}$$

$$E^4(\psi_e)_1 = 3Q_2(\mu)\frac{3k+2}{k+1}\left(\frac{1}{r^2} - \frac{1}{2r^3}\frac{3k+2}{k+1} + \frac{1}{2r^5}\frac{k}{k+1}\right), \tag{A8}$$

where $Q_2(\mu) = -\frac{1}{2}(1-\mu^2)\mu$.

Using the same procedure presented in the previous steps, it can be shown that the stream function equation for the term with the order of $ReCa$ follows:

$$E^4(\psi_j)^{(1)} = 0 \qquad j = i, e. \tag{A9}$$

Here, the stream function is solved by applying the boundary conditions on the new approximated shape. To simplify the problem, Sostarecz and Belmonte [25] recommended a modifications on the boundary condition as:

$$(u_i)_r^{(1)} - \alpha_2 P_2(\cos\theta)\frac{\partial(u_i)_{0,r}}{\partial r} - 3\alpha_2 \cos\theta \sin\theta \, (u_i)_{0,\theta} = 0, \tag{A10}$$

$$(u_e)_r^{(1)} - \alpha_2 P_2(\cos\theta)\frac{\partial(u_e)_{0,r}}{\partial r} - 3\alpha_2 \cos\theta \sin\theta \, (u_e)_{0,\theta} = 0, \tag{A11}$$

$$(u_e)_\theta^{(1)} - \alpha_2 P_2(\cos\theta)\frac{\partial(u_e)_{0,\theta}}{\partial r} + 3\alpha_2 \cos\theta \sin\theta \, (u_e)_{0,r}$$
$$= \frac{1}{(k+1)}\left((u_i)_\theta^{(1)} - \alpha_2 P_2(\cos\theta)\frac{\partial(u_i)_{0,\theta}}{\partial r} + 3\alpha_2 \cos\theta \sin\theta \, (u_i)_{0,r}\right),$$

$$\tag{A12}$$

$$(\tau_e)_{r\theta}^{(1)} - \alpha_2 P_2(\cos\theta)\frac{\partial(\tau_e)_{0,r\theta}}{\partial r} + 3\alpha_2 \cos\theta \sin\theta \left((\tau_e)_{0,rr} - (\tau_e)_{0,\theta\theta}\right)$$
$$= \frac{k}{(k+1)}\left((\tau_i)_{r\theta}^{(1)} - \alpha_2 P_2(\cos\theta)\frac{\partial(\tau_i)_{0,r\theta}}{\partial r} + 3\alpha_2 \cos\theta \sin\theta \left((\tau_i)_{0,rr} - (\tau_i)_{0,\theta\theta}\right)\right).$$

$$\tag{A13}$$

In order to satisfy the normal stress balance boundary condition (Eq. (13)), we must determine $(\tau_i)^{(1)}$ and $(\tau_e)^{(1)}$, and then apply them into the expanded form of Eq. (13) expressed as follows:

$$n.\left((\tau_e)^{(1)} - (\tau_i)^{(1)}\right).n = (\tau_e)_{rr}^{(1)} - \alpha_2 P_2(\cos\theta)\frac{\partial(\tau_e)_{0,rr}}{\partial r} - 6\alpha_2 \cos\theta \sin\theta \, (\tau_e)_{0,r\theta} -$$
$$\frac{k}{(k+1)}\left((\tau_i)_{rr}^{(1)} - \alpha_2 P_2(\cos\theta)\frac{\partial(\tau_i)_{0,rr}}{\partial r} - 6\alpha_2 \cos\theta \sin\theta \, (\tau_i)_{0,r\theta}\right).$$

$$\tag{A14}$$

Subsequently, the first order Deborah term of the stress tensor can be calculated as follows:

$$(\tau_i)_{11} = 2(\boldsymbol{D}_i)_{11} - 2\beta \hat{d}_0 (\boldsymbol{D}_i)_0 - 4\alpha\beta\{(\boldsymbol{D}_i)_0 \cdot (\boldsymbol{D}_i)_0\}. \tag{A15}$$

For a Newtonian fluid, the stress can be written as $(\tau_e)_{11} = 2(\boldsymbol{D}_e)_{11}$. After collecting the first order Deborah terms in the momentum equation, the following equations are obtained:

$$E^4(\psi_i)_{11} = -r\sin(\theta)[\nabla \times \nabla \cdot (\beta \hat{d}_0 (\boldsymbol{D}_i)_0 + \alpha\beta\{(\boldsymbol{D}_i)_0 \cdot (\boldsymbol{D}_i)_0\})], \tag{A16}$$

$$E^4(\psi_e)_{11} = 0. \tag{A17}$$

The details of the perturbation solution for the velocity and pressure fields the order of $De$ are as follows:

$\tau_{i_{rr},11}(r,\theta) = \frac{1}{10k+1}((9.5\alpha r^2 \cos^2\theta \, k - 31.5\alpha r^2 k - 17.5 r^2 \cos^2\theta \, \alpha - 22.5\alpha r^2 - 43 r^2 \cos^2\theta \, k + r^2 k + 65 r^2 \cos^2\theta - 35 r^2 + 36\cos^2\theta \, k - 2k + 10 - 9\alpha\cos^2\theta \, k + 3\alpha k)\beta),$

$\tau_{i_{r\theta},11}(r,\theta) = -\frac{3}{2k+2}(0.6\alpha r^2 k - \alpha r^2 - 2.4 r^2 k - 0.6\alpha k + 4 r^2 + 2.4k)\beta \sin\theta \cos\theta,$

$\tau_{i_{r\phi},11}(r,\theta) = 0,$

$\tau_{i_{\theta\theta},11}(r,\theta) = -\frac{1}{4k+4}((3.4\alpha r^2 \cos^2\theta \, k - 5r^2 \cos^2\theta \, \alpha - 11.6 r^2 \cos^2\theta \, k - 3.6\alpha \cos^2\theta \, k + 22 r^2 \cos^2\theta + 4.2\alpha r^2 k + 14.4\cos^2\theta \, k + 9\alpha r^2 - 8.8 r^2 k + 2.43\alpha k - 28 r^2 - 7.6k + 2)\beta),$

$\tau_{i_{\theta\phi},11}(r,\theta) = 0,$

$\tau_{i_{\phi\phi},11}(r,\theta) = -\frac{1}{2k+2}((3.2\alpha r^2 \cos^2\theta \, k + 2r^2 \cos^2\theta \, \alpha - 5.8 r^2 \cos^2\theta \, k - r^2 \cos^2\theta + 0.6\alpha r^2 k - 4.4 r^2 k - 0.6\alpha k - 2 r^2 + 3.4k + 1)\beta),$ (A18)

$\tau_{e_{rr},11}(r,\theta) = -\frac{1}{r^5(k+1)^2}(0.6(3\alpha r^2 \cos^2\theta - 12 r^2 \cos^2\theta - 3\cos^2\theta \, \alpha - r^2\alpha + 12\cos^2\theta + \alpha - 4)k\beta),$

$\tau_{e_{r\theta},11}(r,\theta) = -\frac{0.9\beta k(\alpha-4)(r^2-1)\sin\theta \cos\theta}{r^5(k+1)^2},$

$\tau_{e_{r\phi},11}(r,\theta) = 0,$

$\tau_{e_{\theta\theta},11}(r,\theta) = -\frac{0.6\beta k(\alpha-4)(2\cos^2\theta - 1)}{r^5(k+1)^2},$

$\tau_{e_{\theta\phi},11}(r,\theta) = 0,$

$\tau_{e_{\phi\phi},11}(r,\theta) = \frac{1.5\beta k(\alpha-4)(0.6\cos^2\theta \, r^2 - 0.2 r^2 - \cos^2\theta + 0.2)}{r^5(k+1)^2}.$ (A19)

The stream functions with the order of $DeCa$ are as follows:

$$(\psi_i)^{(11)} = -\frac{1}{2}\left(\frac{\alpha_2(4-2k)r^4}{5(k+1)} + \frac{(3k-3)r^2}{5(k+1)}\right)\sin(\theta)^2 - \frac{1}{8}\left(\frac{6\alpha_2(5)r^6}{35} + \frac{6\alpha_2(-12)r^4}{35}\right)\sin(\theta)^2(5\cos(\theta)^2 - 1), \tag{A20}$$

$$(\psi_e)^{(11)} = -\frac{1}{2}\left(\frac{\alpha_2}{10(k+1)^2}\left[(3k^2 - k + 8)r + \frac{3k^2 - 3k + 6}{r}\right]\right)\sin(\theta)^2 - \frac{1}{8}\left(\frac{\alpha_2}{35(k+1)}\left[\frac{21k+18}{r} - \frac{21k+4}{r^3}\right]\right)\sin(\theta)^2(5\cos(\theta)^2 - 1). \tag{A21}$$

The pressure with the order of zeroth, *Re*, *ReCa* and *De* for interior and exterior fluids are as follows:

$$(P_i)_0 = \frac{5k\cos(\theta)}{k+1}, \tag{A22}$$

$$(P_i)_1 = \frac{1}{480\,(k+1)^3}(-252k^3 + 20\gamma k - 483k^2 + 20\gamma - 210k)(3\cos(\theta)^2 - 1). \tag{A23}$$

$$(P_i)^{(1)} = -\frac{\alpha_2}{70(k+1)^2}(240\cos(\theta)^3 k + 240\cos(\theta)^3 + 84P_1(\mu)k^2 - 582P_3(\mu)k^2 - 32\cos(\theta)k - 420P_1(\mu)k - 582P_3(\mu)k - 368\cos(\theta)), \tag{A24}$$

$$(P_i)_{11} = \frac{(1.7\cos(\theta)^2\alpha k - 6.1\cos(\theta)^2 k - 5.1\alpha k + 3.8k + 1.3\cos(\theta)^2\alpha - 4.1\alpha + 6.5\cos(\theta)^2 - 8)\beta}{k+1}. \tag{A25}$$

$$(P_e)_0 = \frac{(-1.5k-1)\cos(\theta)}{k+1}, \tag{A26}$$

$$(P_e)_1 = \frac{1}{480\,(k+1)^3}(135k^3 + 333k^2 + 272k + 80)(3\cos(\theta)^2 - 1). \tag{A27}$$

$$(P_e)^{(1)} = -\frac{\alpha_2}{70(k+1)^2}(2520\cos(\theta)^3 k^2 + 2280\cos(\theta)^3 k - 240\cos(\theta)^3 - 1428\cos(\theta)k^2 - 147P_1(\mu)k^2 + 63P_3(\mu)k^2 - 1480\cos(\theta)k + 189P_1(\mu)k - 363P_3(\mu)k + 284\cos(\theta) + 84P_1(\mu) - 426P_3(\mu)), \tag{A28}$$

$$(P_e)_{11} = \frac{3k\beta(3\cos(\theta)^2 - 1)(\alpha - 4)}{10(k+1)^2}. \tag{A29}$$

**Appendix B**

Extensional viscosity, i.e. elongational viscosity, is a viscosity coefficient when applied deformation rate is extensional. This parameter is often used for characterizing polymer solutions. In uniaxial elongational flows, the three components of the velocity vector are $\boldsymbol{u}^* = (\dot{\epsilon}^* x_1^*, -\frac{\dot{\epsilon}^*}{2} x_2^*, -\frac{\dot{\epsilon}^*}{2} x_3^*)$. The rate of elongation in the $x_1$-direction is $\dot{\epsilon}^* = \frac{\partial u_{x_1}^*}{\partial x_1^*}$.

For a Newtonian Fluid, the uniaxial elongational viscosity is three times the shear viscosity ($\eta_{ext}^* = 3\eta^*$) which can be obtained using the continuity equation for incompressible fluids.

In following, responses of the Oldroyd-B, the linearized Giesekus and the full non-linear Giesekus models in uniaxial elongational flows will be investigated.

### 1- Oldroyd-B model:

The stress tensor is consisted of a polymeric contribution and a Newtonian solvent contribution. Polymeric contribution equation is as same as the upper-convected Maxwell model (if the solvent viscosity is zero, the Oldroyd-B becomes the UCM model). The UCM model in tensor form gives six equations:

$$\tau_{xx}^* - 2\lambda^* \dot{\epsilon}^* \tau_{xx}^* = 2\eta_p^* \dot{\epsilon}^*, \tag{B1}$$

$$\tau_{yy}^* + \lambda^* \dot{\epsilon}^* \tau_{yy}^* = -\eta_p^* \dot{\epsilon}^*, \tag{B2}$$

$$\tau_{zz}^* + \lambda^* \dot{\epsilon}^* \tau_{zz}^* = -\eta_p^* \dot{\epsilon}^*, \tag{B3}$$

$$\tau_{xy}^* - \frac{\lambda^* \dot{\epsilon}^*}{2} \tau_{xy}^* = 0, \tag{B4}$$

$$\tau_{xz}^* - \frac{\lambda^* \dot{\epsilon}^*}{2} \tau_{xz}^* = 0, \tag{B5}$$

$$\tau_{yz}^* + \lambda^* \dot{\epsilon}^* \tau_{yz}^* = 0, \tag{B6}$$

From Eqs. B1-B6 we have:

$$\tau_{xx}^* = \frac{2\eta_p^* \dot{\epsilon}^*}{1-2\lambda^* \dot{\epsilon}^*}, \tau_{yy}^* = -\frac{\eta_p^* \dot{\epsilon}^*}{1+\lambda^* \dot{\epsilon}^*}, \tau_{zz}^* = -\frac{\eta_p^* \dot{\epsilon}^*}{1+\lambda^* \dot{\epsilon}^*}, \tau_{xy}^* = \tau_{xz}^* = \tau_{yz}^* = 0, \tag{B7}$$

Since in a Newtonian fluid we have:

$$\tau^*_{xx} = 2\eta^*_s \dot{\epsilon}^*, \tau^*_{yy} = -\eta^*_s \dot{\epsilon}^*, \tau^*_{zz} = -\eta^*_s \dot{\epsilon}^*, \tau^*_{xy} = \tau^*_{xz} = \tau^*_{yz} = 0, \tag{B8}$$

The extensional viscosity of Oldroyd-B fluid will be:

$$\eta^*_{ext} = \frac{\tau^*_{xx} - \tau^*_{yy}}{\dot{\epsilon}} = 3\eta^*_s + \frac{3\eta^*_p}{(1-2\lambda^*\dot{\epsilon}^*)(1+\lambda^*\dot{\epsilon}^*)}, \tag{B9}$$

Figure B1 illustrates extensional viscosity of the Oldroyd-B model in steady extensional uniaxial flow at $\beta = 0.9$. It is seen that the extensional viscosity increases by increasing strain rate for low strain rates (see in inset). For higher strain rates, the extensional viscosity diverges at $\lambda^*\dot{\epsilon}^* = 0.5$. This leads to negative viscosities at larger strain rates (Fig. B1) which is "unphysical".

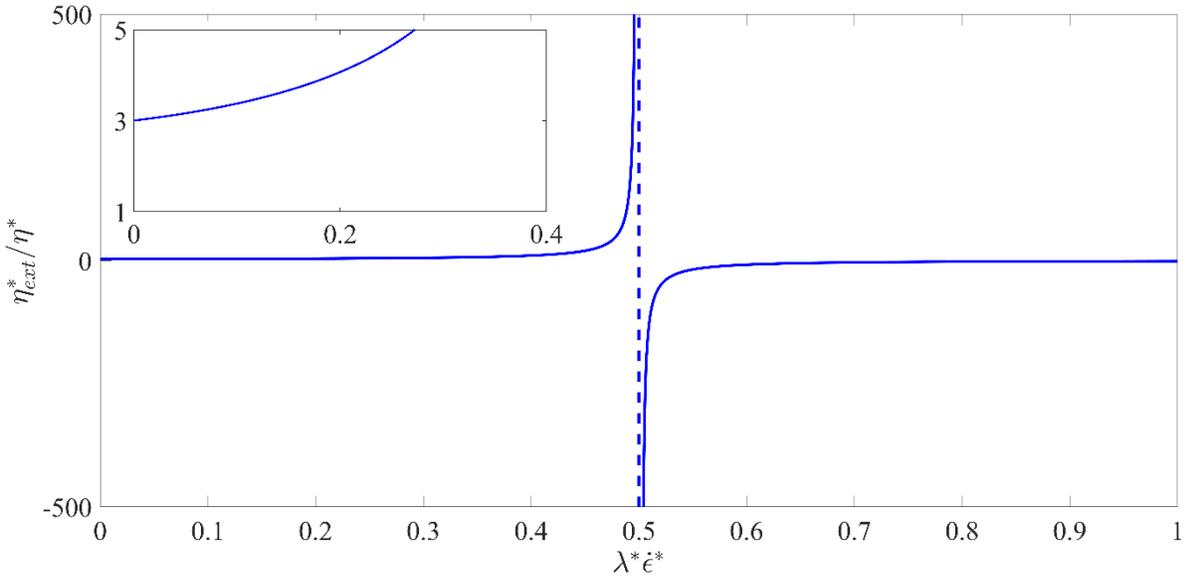

**Fig. B1:** Extensional viscosity of the Oldroyd-B model in steady extensional uniaxial flow scaled by the total shear viscosity for $\beta = 0.9$.

2- **The linearized Giesekus model:**

Using the Giesekus model (Eq. (3b)) and perturbation method, the first order Deborah term of the stress tensor can be calculated as follows Eq. (A15) (The linearized Giesekus model). Eq. (A15) in tensor form gives six equations:

$$\tau_{xx}^* - 2\lambda^*\dot{\epsilon}^*\tau_{xx}^* = 2\eta_{p,0}^*\dot{\epsilon}^*(1 - 2\alpha\lambda^*\dot{\epsilon}^*), \tag{B10}$$

$$\tau_{yy}^* + \lambda^*\dot{\epsilon}^*\tau_{yy}^* = -\eta_{p,0}^*\dot{\epsilon}^*(1 + \alpha\lambda^*\dot{\epsilon}^*), \tag{B11}$$

$$\tau_{zz}^* + \lambda^*\dot{\epsilon}^*\tau_{zz}^* = -\eta_{p,0}^*\dot{\epsilon}^*(1 + \alpha\lambda^*\dot{\epsilon}^*), \tag{B12}$$

$$\tau_{xy}^* - \frac{\lambda^*\dot{\epsilon}^*}{2}\tau_{xy}^* = 0, \tag{B13}$$

$$\tau_{xz}^* - \frac{\lambda^*\dot{\epsilon}^*}{2}\tau_{xz}^* = 0, \tag{B14}$$

$$\tau_{yz}^* + \lambda^*\dot{\epsilon}^*\tau_{yz}^* = 0, \tag{B15}$$

From Eqs. B10-B15 can obtain:

$$\tau_{xx}^* = \frac{2\eta_{p,0}^*\dot{\epsilon}^*(1-2\alpha\lambda^*\dot{\epsilon}^*)}{1-2\lambda^*\dot{\epsilon}^*}, \tau_{yy}^* = -\frac{\eta_{p,0}^*\dot{\epsilon}^*(1+\alpha\lambda^*\dot{\epsilon}^*)}{1+\lambda^*\dot{\epsilon}^*}, \tau_{zz}^* = -\frac{\eta_{p,0}^*\dot{\epsilon}^*(1+\alpha\lambda^*\dot{\epsilon}^*)}{1+\lambda^*\dot{\epsilon}^*},$$

$$\tau_{xy}^* = \tau_{xz}^* = \tau_{yz}^* = 0, \tag{B16}$$

In a Newtonian fluid we have in Eq. (B8). The linear Giesekus fluid has extensional viscosity:

$$\eta_{ext}^* = 3\eta_s^* + \frac{2\eta_{p,0}^*(1-2\alpha\lambda^*\dot{\epsilon}^*)(1+\lambda^*\dot{\epsilon}^*) + \eta_{p,0}^*(1+\alpha\lambda^*\dot{\epsilon}^*)(1-2\lambda^*\dot{\epsilon}^*)}{(1-2\lambda^*\dot{\epsilon}^*)(1+\lambda^*\dot{\epsilon}^*)},$$

Or (B17)

$$\eta_{ext}^* = 3\eta_s^* + \frac{2\eta_{p,0}^*(1-2\alpha\lambda^*\dot{\epsilon}^*) + \eta_{p,0}^*(1+\alpha\lambda^*\dot{\epsilon}^*) - 6\eta_{p,0}^*\alpha\lambda^{*2}\dot{\epsilon}^{*2}}{(1-2\lambda^*\dot{\epsilon}^*)(1+\lambda^*\dot{\epsilon}^*)},$$

Figure B2 depicts extensional viscosity of the linearized Giesekus model in steady extensional uniaxial flow for different mobility factors ($\alpha$) at $\beta = 0.9$. The mobility factor controls the level of "non-linearity" in full Giesekus model before linearization. As shown, the extensional viscosity is reduced by increasing the mobility factor ($\alpha$). A simple comparison between equations B1-B3 and B10-B12 shows that applying a perturbation method in elongational dominated flows can "transform" the Giesekus model to a modified version of the Oldroyd-B model and so at $\lambda^*\dot{\epsilon}^* = 0.5$, the extensional viscosity diverges again (singularity point).

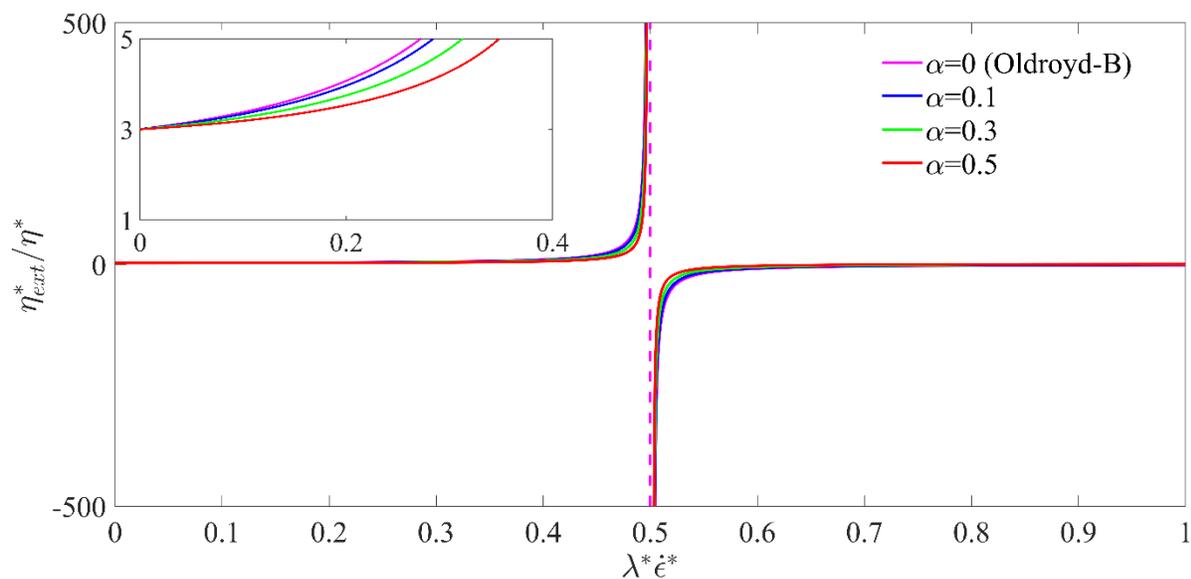

**Fig. B2:** Extensional viscosity of the linear Giesekus model in steady extensional uniaxial flow for different mobility factors ($\alpha$) at $\beta = 0.9$.

### 3- The full non-linear Giesekus model:

Using the full non-linear Giesekus model (Eq. (3b)), the polymeric conurbation of the Giesekus model in tensor form gives six equations:

$$\tau_{xx}^* - 2\lambda^*\dot{\epsilon}^*\tau_{xx}^* + \frac{\alpha\lambda^*}{\eta_{p,0}^*}(\tau_{xx}^{*2} + \tau_{xy}^{*2} + \tau_{xz}^{*2}) = 2\eta_{p,0}^*\dot{\epsilon}^*, \tag{B18}$$

$$\tau_{yy}^* + \lambda^*\dot{\epsilon}^*\tau_{yy}^* + \frac{\alpha\lambda^*}{\eta_{p,0}^*}(\tau_{xy}^{*2} + \tau_{yy}^{*2} + \tau_{yz}^{*2}) = -\eta_{p,0}^*\dot{\epsilon}^*, \tag{B19}$$

$$\tau_{zz}^* + \lambda^*\dot{\epsilon}^*\tau_{zz}^* + \frac{\alpha\lambda^*}{\eta_{p,0}^*}(\tau_{xz}^{*2} + \tau_{yz}^{*2} + \tau_{zz}^{*2}) = -\eta_{p,0}^*\dot{\epsilon}^*, \tag{B20}$$

$$\tau_{xy}^* - \frac{\lambda^*\dot{\epsilon}^*}{2}\tau_{xy}^* + \frac{\alpha\lambda^*}{\eta_{p,0}^*}(\tau_{xx}^*\tau_{xy}^* + \tau_{xy}^*\tau_{yy}^* + \tau_{xz}^*\tau_{yz}^*) = 0, \tag{B21}$$

$$\tau_{xz}^* - \frac{\lambda^*\dot{\epsilon}^*}{2}\tau_{xz}^* + \frac{\alpha\lambda^*}{\eta_{p,0}^*}(\tau_{xx}^*\tau_{xz}^* + \tau_{xy}^*\tau_{yz}^* + \tau_{xz}^*\tau_{zz}^*) = 0, \tag{B22}$$

$$\tau_{yz}^* + \lambda^*\dot{\epsilon}^*\tau_{yz}^* + \frac{\alpha\lambda^*}{\eta_{p,0}^*}(\tau_{xy}^*\tau_{xz}^* + \tau_{yy}^*\tau_{yz}^* + \tau_{yz}^*\tau_{zz}^*) = 0, \tag{B23}$$

From Eqs. B18-B23 can obtain:

$$\tau_{xx}^* = \frac{\eta_{p,0}^*\left(2\lambda^*\dot{\epsilon}^* - 1 + \sqrt{4\lambda^{*2}\dot{\epsilon}^{*2} + 4\lambda^*\dot{\epsilon}^*(2\alpha-1) + 1}\right)}{2\alpha\lambda^*}, \tau_{yy}^* = \frac{\eta_{p,0}^*\left(-\lambda^*\dot{\epsilon}^* - 1 + \sqrt{\lambda^{*2}\dot{\epsilon}^{*2} - 2\lambda^*\dot{\epsilon}^*(2\alpha-1) + 1}\right)}{2\alpha\lambda^*},$$

$$\tau_{zz}^* = \frac{\eta_{p,0}^*\left(-\lambda^*\dot{\epsilon}^* - 1 + \sqrt{\lambda^{*2}\dot{\epsilon}^{*2} - 2\lambda^*\dot{\epsilon}^*(2\alpha-1) + 1}\right)}{2\alpha\lambda^*}, \tau_{xy}^* = \tau_{xz}^* = \tau_{yz}^* = 0, \tag{B24}$$

In a Newtonian fluid we have in Eq. (B8). The full non-linear Giesekus fluid has extensional viscosity:

$$\eta_{ext}^* = 3\eta_s^* + \frac{3\eta_{p,0}^*\lambda^*\dot{\epsilon}^* + \eta_{p,0}^*\left(\sqrt{4\lambda^{*2}\dot{\epsilon}^{*2} + 4\lambda^*\dot{\epsilon}^*(2\alpha-1) + 1} - \sqrt{\lambda^{*2}\dot{\epsilon}^{*2} - 2\lambda^*\dot{\epsilon}^*(2\alpha-1) + 1}\right)}{2\alpha\lambda^*\dot{\epsilon}^*}, \tag{B25}$$

A comparison between the extensional viscosity of the linear Giesekus model and the full non-linear Giesekus model for different mobility factors ($\alpha$) at $\beta = 0.9$ is presented in Fig. B3.

Unlike the linear Giesekus model (as same as Oldroyd-B model) which predicts a singularity followed by non-physical negative values of the elongational viscosity after a critical strain rate, in the full non-linear Giesekus model the elongational viscosity varies between two bounded values. In the full non-linear Giesekus model, by increasing mobility factors ($\alpha$) value of the elongational viscosity reduces at the strain rate fixed. The results suggest that linearization of the model have a good agreement for small values of $\lambda^* \dot{\epsilon}^*$ when its values is less than half.

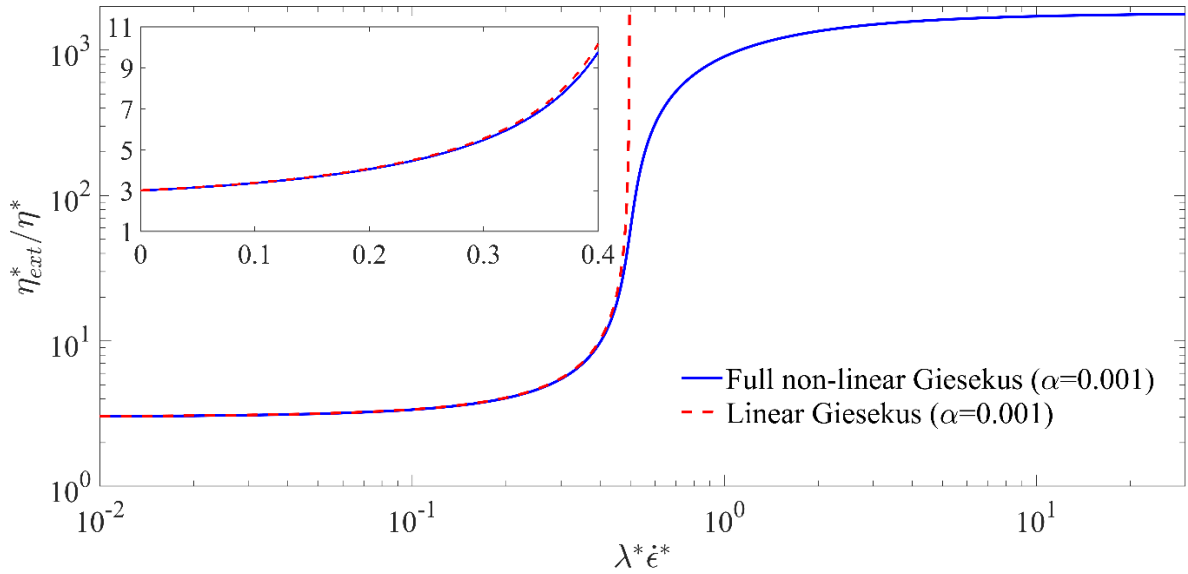

**Fig. B3:** Comparison between extensional viscosity of the linear Giesekus model and the full non-linear Giesekus model for different mobility factors at $\beta = 0.9$.